\let\new=\newcommand
\new{\eq}{\begin{equation}}
\new{\en}{\end{equation}}
\new{\rzim}{$r_{0}$}
\new{\vtwo}{$v_{200}$}
\new{\cn}{$C^2_0$}

\documentclass[letterpaper,12pt]{article}   %% LaTeX 2e (preferred)
\usepackage{opex3} %% do not use with REVTeX4

\input epsf
 
\begin{document}

\title{Improved models of upper-level wind for several astronomical observatories}

\author{Lewis C. Roberts, Jr$^{1,*}$ and L. William Bradford$^{2}$}

\address{$^1$Jet Propulsion Laboratory, California Institute of Technology \\4800 Oak Grove Dr, Pasadena CA 91009, USA}
\address{$^2$Pacific Defense Solutions, \\ 1300 N. Holopono St. Suite 116, Kihei, HI 96753, USA}
\begin{center} \textit{*lewis.c.roberts@jpl.nasa.gov} \end{center}

%\maketitle %% null function with osajnl.sty

Accepted to Optics Express
%%%%%%%%%%%%%%%%%%%%%%%%%%%%%%%%%%%%%%%%%%%%%%%%%%%%%%%%%%%%%%%

\begin{abstract}
An understanding of wind speed and direction as a function of height are critical to the proper modeling of atmospheric turbulence.   We have used radiosonde data from launch sites near significant astronomical observatories and created mean profiles of wind speed and direction and have also computed Richardson number profiles.  Using data from the last 30 years, we extend the  1977 Greenwood wind profile  to include parameters that show seasonal variations and differences in location. The added information from our models is useful for the design of adaptive optics systems and other imaging systems. Our analysis of the Richardson number suggests that persistent turbulent layers may be inferred when low values  are present in our long term averaged data. Knowledge of the presence of these layers may help with planning for adaptive optics and laser communications.
\end{abstract}
 

%%%%%%%%%%%%%%%%%%%%%%%%%%%%%%%%%%%%%%%%%%%%%%%%%%%%%%%%%%%%%%%

% References

%%%%%%%%%%%%%%%%%%%%%%%%%%%%%%%%%%%%%%%%%%%%%%%%%%%%%%%%%%%%%%%

\section{Introduction}  

Images collected with ground based telescopes show temporal variation in image quality, primarily due to fluctuations in the index of refraction of the air along the light path.  These index of refraction fluctuations are caused by differences in air temperature arising from turbulent mixing. These turbulent motions happen on smaller scales than the gross motions of the winds. Still it is known that the wind has a part in the strength of these fluctuations.  Hufnagel \cite{hufnagel1974} derived a model for the vertical profile of strength of turbulence $C_n^2(z)$ which depended on a wind speed profile centered at a particular altitude. Furthermore, it has been observed that the index of refraction fluctuations are enhanced where there are shears in winds \cite{beland1993}, so that direction as well as speed can be a factor, although this is difficult to quantify \cite{dewan1993}.

Aside from the effects of wind on turbulence strength, the movement of turbulent regions past an observer induces temporal variations in seeing. If wind speeds are sufficiently high that turbulence does not evolve significantly over a measurement period, we apply Taylor's hypothesis and treat the temperature fluctuations as quantities that are advected by the wind \cite{taylor1938}. 

The following wind model is often used in the modeling of adaptive optics and other imaging systems (\textit{e.g.} \cite{parenti1994, andrews1998,tyson1998}),

\begin{equation}
\label{greenwood_fit}
   v(z) = 8 + 30 \exp\left(-\left[\frac{(z\sin\theta -9400)}{4800}\right]^2\right),
\end{equation}

\noindent where $v(z)$ is the wind speed in m/s, $z$ is the height in m, and $\theta$ is the angle from the zenith. Bufton \cite{bufton1973} is often cited as the source for this model (\textit{e.g.} \cite{parenti1994, andrews1998,tyson1998}), but that paper does not present an explicit wind model. 

The basic form of the model was presented by Greenwood \cite{greenwood1977}.  Readers often over look the statement in the body of the original paper that the height $z=0$ corresponds to a mean sea level altitude of 3048 m.  This is the height of the Maui Space Surveillance System on Haleakala, where the analysis was being used.  So the height of the tropopause term, 9400 m, should really be 12448 m when the model is used for sites other than Haleakala.   This caveat has been overlooked in a number of references, leading to misleading results.

The model from Greenwood \cite{greenwood1977} was derived by averaging radiosonde data collected from balloons launched from Lihue on the island of Kauai from 1950-1970 and from Hilo on the island of Hawaii from 1950-1974.  This was more fully detailed in a technical report, which did not explicitly generalize the equation, but the authors clearly understood what the various terms signified  \cite{greenwood1975}.

The wind model is generalized as,

\begin{equation}
\label{gaussian_fit}
% v(z)=v_G + v_T\exp\left(-\left[\frac{z\cos\zeta-H_T}{L_T}\right]^2\right)\left(\sin^2\phi + \cos^2\phi\cos^2\zeta\right)^2
v(z)=v_G + v_T\exp\left(-\left[\frac{z\cos\zeta-h_T}{L_T}\right]^2\right)
\end{equation}

\noindent where $v$ is the wind speed, $z$ is the altitude,  $v_G$ is the wind speed at the ground or low altitude, $v_T$ is the wind speed at tropopause, $h_T$ is the altitude of the tropopause, $L_T$ is the thickness of the tropopause layer and $\zeta$ is the zenith angle of the observation \cite{hardy1998}.

There are several uses for a wind profile.  It can be used in conjunction with a turbulence model to predict the performance of instrumentation.  Most commonly this is applied to the servo bandwidth of closed loop systems  \cite{greenwood1977}.  The minimum bandwidth of the system (to limit servo lag) is related to the Greenwood frequency. Because the Greenwood frequency depends on a product of $C_n^2$ and a power of the wind speed $v^{5/3}$, it is important to get the wind profile and turbulence profiles as a function of altitude.  A wind profile can also be used to simulate turbulence in conjunction with phase screens.

Currently, there is an increased emphasis on modeling effects of turbulence on telescopes with diameters of 20-50 meters. Significantly more spatial variation of wavefront optical path difference is captured with such large telescopes in a short exposure than on a smaller telescope. Some of that variation is due to the presence of multiple layers of turbulence, moving at different speeds and in different directions. The patterns may not simply translate across the telescope aperture as implied by the Taylor hypothesis, but are more likely to``boil" with a complicated wind profile. This can have an impact on the design and performance of an adaptive optics system.   

In order to better model atmospheric turbulence, we used archival radiosonde data to compute the wind profiles for a number of astronomical observatories.  We also show that the wind profile has a strong seasonal variation. We also extend the wind profiles, by including wind direction. Unfortunately, the wind directions do not always lend themselves to an analytic expression, so we only show them in graphs.

Using the same radiosonde data, we examined the variation of the gradient Richardson number \cite{kundu1990}), which is an indicator of the stability of parcels of air. Since it depends upon strength of wind shear, as opposed to wind speed, it also serves to inform the modeler that wind speed alone is not sufficient. As will be seen in our plots, the wind speed can be modeled with a few simple terms, but the form of the Richardson number plots indicates the true nature of atmospheric turbulence: layered, with sharp boundaries (although these are muted by the resolution of our data).

\section{Data Analysis}\label{data_analysis}

Worldwide there are over 900 sites that launch radiosondes on a routine basis. Radiosondes are small instrument packages mounted on weather balloons.  They record atmospheric pressure, temperature, potential temperature, mixing ratio, dew point, and relative humidity as a function of height. Data are taken at about 6 second intervals, but databases may only have the data recorded at 60 second intervals. Readings at certain specific standard barometric pressures are made regardless of when they occur.  Since the balloons rise at about 5 m/s, the best vertical resolution would be approximately
30 meters, while the usual recorded resolution would be 300 meters and frequently the recorded data are at irregular intervals.  Wind speed and wind direction are obtained from either radar observation or a navigation system such as LOng RAnge Navigation (LORAN) and newer systems may use the Global Positioning System. Almost all of these sites launch balloons at 0 UT and 12 UT. The information is used as entries to weather prediction and simulation programs. 

% Further information can be found at the National Weather Service web site (http://www.ua.nws.noaa.gov) for upper air measurements, and the links to other web pages there.

We downloaded atmospheric sounding data for 1973 January till 2006 September for a number of stations from the University of Wyoming's Weather Web (http://weather.uwyo.edu/upperair/sounding.html).  The stations were chosen based upon their proximity to an astronomical observatory.   Not all the sites had data for the entire time period; the dates for the available data are also listed in the table. In addition there are gaps of varying length in each data set, which are most likely due to equipment problems.  These data drop outs are not a significant problem due to the large number of data points that are available.  The names of stations and their details along with the nearby observatories are listed in Table \ref{station_table}. The name of the station is given, along with its latitude, longitude and altitude, the dates when data were recorded and the names of nearby observatories. In parenthesis are the distance in km between the observatory and the radiosonde launch site and the altitude of the observatory in km.

%\begin{table}[htb]
%{\bf \caption{\label{station_table}Details of the radiosonde launch sites used}}
%\begin{center}
%\begin{tabular}{lccccl}
%\hline
%Station     & Lat.         & Lon.        & Alt.& Dates & Nearby \\
%            &   ($^\circ$) &  ($^\circ$) &  (m) &  & Observatories\\
%\hline
%Antofagasta, Chile & -23.43  & -70.43    &    135    & 1973-2006      & Paranal,\\
%Antofagasta, Chile & -23.43  &   &         &     &  Armazones\\
%Flagstaff, AZ      &  35.23  &  -111.82  &   2192    & 1995-2006      & Lowell\\
%Hilo, HI           &  19.71  &   -155.06  &   11     & 1973-2006      & Haleakala, Mauna Kea\\
%Oakland, CA        &  37.72  &   -122.20  &   3      & 1973-2006      & Lick\\
%San Diego, CA      &  32.84  &   -117.11  &   128    & 1990-2006      & Palomar\\
%Tenerife, Spain    &  28.47  &   -16.38   &   105    & 2002-2006      & R. Muchachos, Teide \\
%Tucson, AZ         &  32.11  &   -110.93  &   779    & 1973-2006      & Kitt Peak, Mt. Graham\\
%\hline
%\end{tabular}
%\end{center}
%\end{table}

\begin{table}[htb]
{\bf \caption{\label{station_table}Details of the radiosonde launch sites used}}
\begin{center}
\begin{tabular}{lcccclcc}
\hline
Station     & Lat.         & Lon.        & Alt.& Dates & Nearby & Dist. & Alt.\\
            &   ($^\circ$) &  ($^\circ$) &  (m) &  & Observatories& (km) & (km)\\
\hline
Antofagasta & -23.43  & -70.43    &    135    & 1973-2006      & Paranal & 133  &  2.3  \\
  &   &   &         &     &                                     Armazones &132 &  3.0 \\
Flagstaff     &  35.23  &  -111.82  &   2192    & 1995-2006      & Lowell &14  &  2.2 \\
Hilo           &  19.71  &   -155.06  &   11     & 1973-2006      & Haleakala &168 &   3.1 \\
  &   &   &         &     &                                       Mauna Kea &46  &  4.4\\
Oakland        &  37.72  &   -122.20  &   3      & 1973-2006      & Lick &65  &  1.3 \\
San Diego     &  32.84  &   -117.11  &   128    & 1990-2006      & Palomar &61  &  1.7 \\
Tenerife    &  28.47  &   -16.38   &   105    & 2002-2006      & R. Muchachos &150 &  2.3  \\
 &   &   &         &     &                                         Teide &23  &  2.4 \\
Tucson         &  32.11  &   -110.93  &   779    & 1973-2006      & Kitt Peak &64  &  2.1 \\
 &   &   &         &     &                                         Mt. Graham &117 &  3.1\\

%Antofagasta & -23.43  & -70.43    &    135    & 1973-2006      & Paranal (133km, 2.3km) \\
%  &   &   &         &     &  Armazones (132km, 3.0 km) \\
%Flagstaff     &  35.23  &  -111.82  &   2192    & 1995-2006      & Lowell (14 km, 2.2 km) \\
%Hilo           &  19.71  &   -155.06  &   11     & 1973-2006      & Haleakala (168 km, 3.1 km) \\
%  &   &   &         &     &   Mauna Kea (46 km, 4.4 km)\\
%Oakland        &  37.72  &   -122.20  &   3      & 1973-2006      & Lick (65 km, 1.3 km) \\
%San Diego     &  32.84  &   -117.11  &   128    & 1990-2006      & Palomar (61 km, 1.7 km) \\
%Tenerife    &  28.47  &   -16.38   &   105    & 2002-2006      & R. Muchachos (150 km, 2.3 km)  \\
% &   &   &         &     &   Teide (23 km, 2.4 km) \\
%Tucson         &  32.11  &   -110.93  &   779    & 1973-2006      & Kitt Peak (64 km, 2.1 km) \\
% &   &   &         &     &   Mt. Graham (117 km, 3.1 km)\\
\hline
\end{tabular}
\end{center}
\end{table}

For each sounding the pressure, temperature, dew point, relative humidity, wind speed and direction are measured as a function of height.   The heights are irregularly measured and not repeated, so the data were interpolated onto a fixed height grid with an increment of 400 m and the mean was computed for various time periods.  The height grid is measured from mean sea level, rather than the local surface.  The different soundings have data measured to varying maximum altitudes, depending on when the balloons burst.  We did not analyze any data from altitudes higher than 30 km, since there were too few data points for a meaningful analysis. 

\begin{figure}[htbp]
\centering\includegraphics[height=4.5cm]{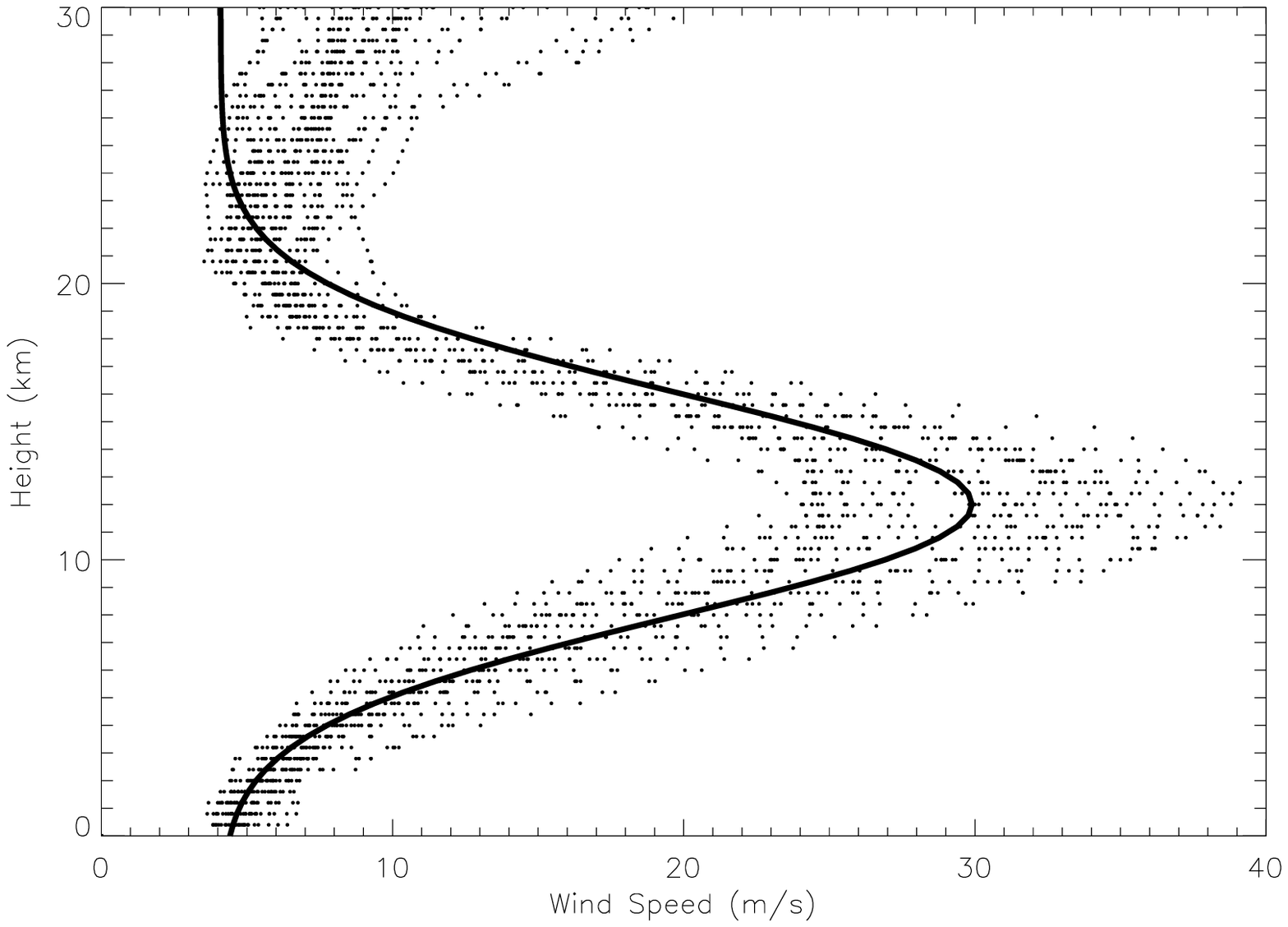}
\includegraphics[height=4.5cm]{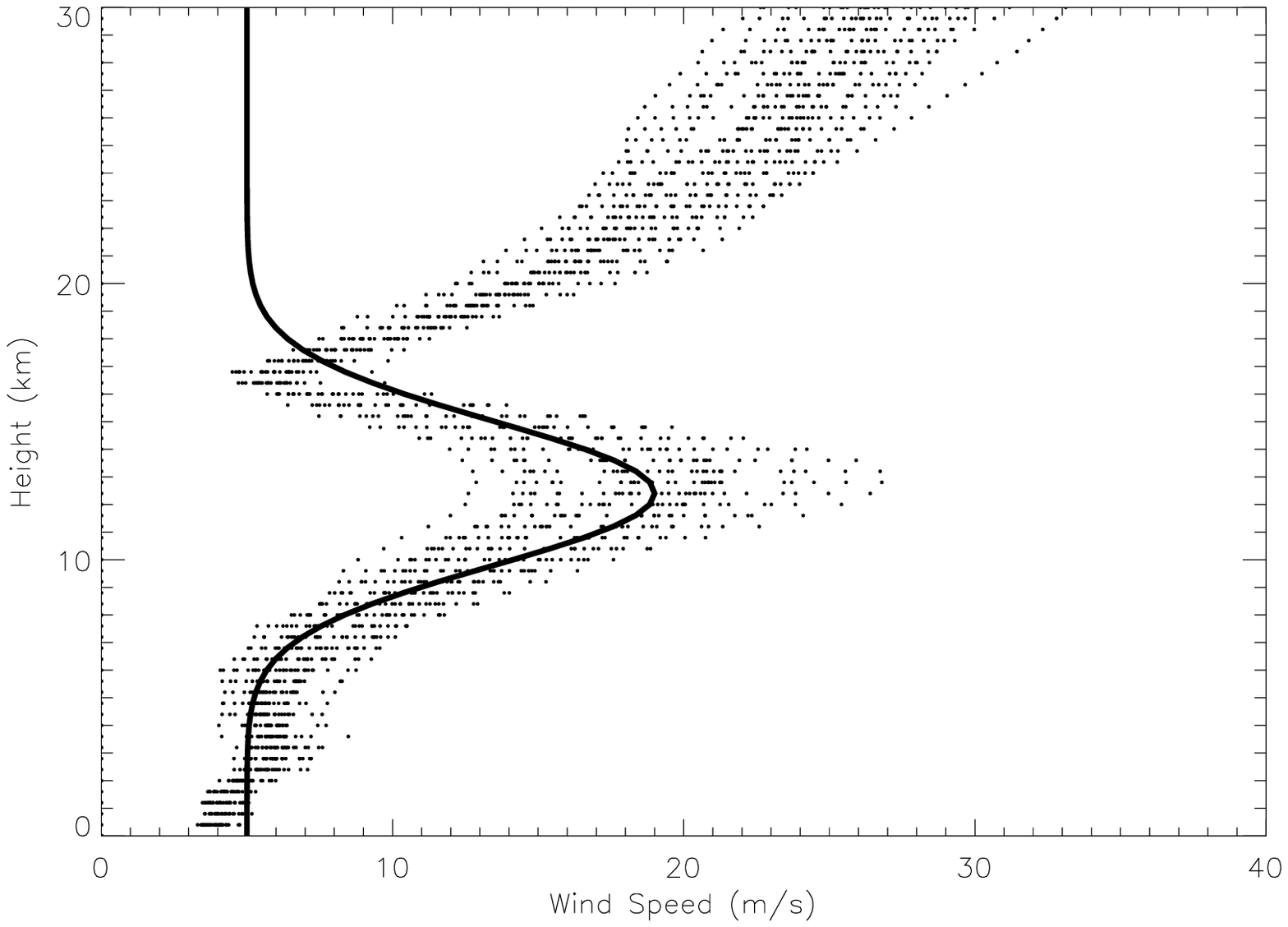}
\caption{\label{speed_variations}   The solid line in the figure on the left is the model fit to the mean monthly wind speed for Hilo, Hawaii in the month of January. The dotted lines are the mean wind speed for Hilo, Hawaii in each January from 1973-2006.  The figure on the right is the same, but for July.  These figures illustrate the variation in the monthly mean wind speed.}
\end{figure}

Fig. \ref{speed_variations} shows the yearly variations that occur for January and July in Hilo, Hawaii.   Plotted over this are the mean wind speeds for the respective month for the entire 33 year period of data. This shows the variation in monthly mean wind speeds.  Some of this difference is caused by large scale weather patterns and some is the natural chaotic behavior of weather. The overall shape of the wind profiles is highly consistent over the 33 years of data.   

In addition to the sites listed in Table \ref{station_table}, we also downloaded the wind profiles for Lihue HI.  As mentioned above this was used in Greenwood's original model.  We computed the correlation coefficient between the Hilo and Lihue data sets for each month of the mean wind profiles.  The correlation was highest in February (0.999) and lowest in August (0.981).  The average correlation was 0.992 with a standard deviation of 0.006.  This correlation was high enough that we felt it was valid to only use the Hilo data.  This does not say that Hilo and Lihue constantly experience the exact same wind conditions, but that on average the wind profiles are very similar. This also shows that it is valid to use the same wind model for the entire state of Hawaii including the observatories on Haleakala and Mauna Kea. 

Surface winds are highly dependent on local geography, such as being in the lee of a mountain, and are also time dependent due to on-shore and off-shore breezes for the island sites and katabatic flows for sites near mountains.  They are also dependent on the solar heating of the ground.    This is less of a problem for astronomical uses of the wind profiles since mountain top observatories are usually situated above these surface effects. In addition, the wind profiles do not capture the effect that the local orography will have on the wind profile.

We created the mean wind speed and direction for each of the 12 months for each of the sites listed in Table \ref{station_table}.  We used a non-linear least squares fit algorithm to fit the data to

\begin{equation}
\label{gaussian_fit_eqn}
   v(z) = A_0 + A_1 \exp\left(-\left[\frac{(z-A_2)}{A_3}\right]^2\right).
\end{equation}

\noindent Tables \ref{hilo_coeffs}-\ref{anto_coeffs} list the monthly coefficients and their associated error bars for all sites listed in Table \ref{station_table}.  The error bars are computed by the non-linear least squares fit.  The errors bars increase as the winds vary from the pure Gaussian model. 
We fit the Gaussian to the wind speed data from the ground to a height of 16 km, beyond this the stratospheric winds start to affect the fit.  The effect is marginal and often below the error bars, except for the summer data sets.

 The Gaussian wind model of \cite{greenwood1977} (Eq. \ref{greenwood_fit}) only fit the troposphere winds and ignored the stratospheric winds.  It is possible to fit both of those winds if a Gaussian is summed with a second order polynomial.  This has limited utility to astronomical imaging, as the turbulence above 20km is relatively weak, but high altitude airships are being designed to operate between 20 and 30 km \cite{fesen2006} and require  a knowledge of the wind speeds and directions in this region.

%Hawaii Wind Speed Model
\begin{table}[htbp]
{\bf \caption{\label{hilo_coeffs}Hawaii Wind Speed Model Coefficients}}
\begin{center}
\begin{tabular}{lcccc}
\hline
Month& A0 & A1 & A2 & A3 \\
\hline
Jan. &  4.1 $\pm$  0.4 &    25.8 $\pm$  0.5 &    12007 $\pm$   69 &    4047 $\pm$ 115\\
Feb. &  4.2 $\pm$  0.4 &    27.6 $\pm$  0.4 &    11952 $\pm$   61 &    3957 $\pm$  99\\
Mar. &  5.0 $\pm$  0.3 &    29.6 $\pm$  0.4 &    12335 $\pm$   55 &    3405 $\pm$  79\\
Apr. &  5.0 $\pm$  0.3 &    29.4 $\pm$  0.4 &    12776 $\pm$   51 &    3001 $\pm$  68\\
May. &  4.8 $\pm$  0.4 &    22.9 $\pm$  0.5 &    12909 $\pm$   88 &    3161 $\pm$ 117\\
Jun. &  4.8 $\pm$  0.3 &    17.0 $\pm$  0.5 &    12637 $\pm$   95 &    2877 $\pm$ 127\\
Jul. &  5.0 $\pm$  0.3 &    14.0 $\pm$  0.5 &    12406 $\pm$  102 &    2605 $\pm$ 133\\
Aug. &  4.6 $\pm$  0.3 &    12.8 $\pm$  0.5 &    12527 $\pm$   99 &    2575 $\pm$ 128\\
Sep. &  4.2 $\pm$  0.3 &    13.6 $\pm$  0.4 &    12263 $\pm$   93 &    2868 $\pm$ 126\\
Oct. &  4.5 $\pm$  0.3 &    17.2 $\pm$  0.4 &    12222 $\pm$   75 &    2845 $\pm$ 101\\
Nov. &  5.1 $\pm$  0.3 &    17.4 $\pm$  0.5 &    12313 $\pm$   92 &    3214 $\pm$ 129\\
Dec. &  5.0 $\pm$  0.4 &    21.1 $\pm$  0.5 &    12199 $\pm$   84 &    3617 $\pm$ 126\\
\hline
\end{tabular}
\end{center}
\end{table}

%Oakland Wind Speed Model
\begin{table}[htbp]
{\bf \caption{\label{oakland_coeffs}Oakland Wind Speed Model Coefficients}}
\begin{center}
\begin{tabular}{lcccc}
\hline
Month& A0 & A1 & A2 & A3\\
\hline
Jan. & -1.4 $\pm$  1.8 &    30.5 $\pm$  1.7 &    10730 $\pm$   81 &    8329 $\pm$ 318\\
Feb. & -0.7 $\pm$  1.6 &    30.0 $\pm$  1.5 &    10695 $\pm$   77 &    8047 $\pm$ 283\\
Mar. & -0.9 $\pm$  1.4 &    29.5 $\pm$  1.3 &    10548 $\pm$   66 &    7878 $\pm$ 245\\
Apr. &  0.6 $\pm$  1.0 &    28.3 $\pm$  1.0 &    10472 $\pm$   59 &    7195 $\pm$ 188\\
May. &  2.8 $\pm$  0.7 &    22.8 $\pm$  0.7 &    10649 $\pm$   70 &    6367 $\pm$ 170\\
Jun. &  3.1 $\pm$  0.7 &    19.7 $\pm$  0.7 &    11164 $\pm$   98 &    6430 $\pm$ 215\\
Jul. &  4.0 $\pm$  0.6 &    16.5 $\pm$  0.6 &    11581 $\pm$  122 &    5540 $\pm$ 210\\
Aug. &  4.4 $\pm$  0.5 &    17.3 $\pm$  0.6 &    11602 $\pm$  113 &    5187 $\pm$ 184\\
Sep. &  3.2 $\pm$  0.6 &    18.8 $\pm$  0.6 &    11400 $\pm$   95 &    6240 $\pm$ 192\\
Oct. &  1.3 $\pm$  0.8 &    22.3 $\pm$  0.8 &    11073 $\pm$   79 &    7297 $\pm$ 217\\
Nov. & -2.0 $\pm$  1.6 &    32.8 $\pm$  1.5 &    10878 $\pm$   72 &    8251 $\pm$ 263\\
Dec. & -1.8 $\pm$  2.0 &    31.1 $\pm$  1.8 &    10611 $\pm$   81 &    8358 $\pm$ 333\\
\hline
\end{tabular}
\end{center}
\end{table}

%San Diego Wind Speed Model
\begin{table}[htbp]
{\bf \caption{\label{sandiego_coeffs}San Diego  Wind Speed Model Coefficients}}
\begin{center}
\begin{tabular}{lcccc}
\hline
Month& A0 & A1 & A2 & A3\\
\hline
 Jan. & -2.9 $\pm$  1.4 &    33.9 $\pm$  1.3 &    11470 $\pm$   76 &    6104 $\pm$ 249\\
 Feb. &  1.4 $\pm$  1.1 &    32.4 $\pm$  1.0 &    11400 $\pm$   81 &    5081 $\pm$ 199\\
 Mar. & -1.3 $\pm$  1.2 &    32.7 $\pm$  1.1 &    11701 $\pm$   82 &    5793 $\pm$ 226\\
 Apr. & -2.4 $\pm$  1.4 &    32.7 $\pm$  1.3 &    11656 $\pm$   88 &    6062 $\pm$ 266\\
 May. &  0.2 $\pm$  1.2 &    25.6 $\pm$  1.1 &    11577 $\pm$  113 &    5471 $\pm$ 295\\
 Jun. &  3.6 $\pm$  0.9 &    17.6 $\pm$  0.9 &    12081 $\pm$  201 &    4546 $\pm$ 364\\
 Jul. &  3.7 $\pm$  0.6 &    10.6 $\pm$  0.6 &    11483 $\pm$  180 &    3909 $\pm$ 317\\
 Aug. &  4.4 $\pm$  0.4 &    10.5 $\pm$  0.5 &    11821 $\pm$  161 &    3318 $\pm$ 239\\
 Sep. &  4.4 $\pm$  0.6 &    18.8 $\pm$  0.6 &    11897 $\pm$  121 &    3745 $\pm$ 193\\
 Oct. &  2.2 $\pm$  0.8 &    22.7 $\pm$  0.7 &    11805 $\pm$  111 &    4843 $\pm$ 230\\
 Nov. & -1.6 $\pm$  1.3 &    30.6 $\pm$  1.2 &    11511 $\pm$   88 &    5776 $\pm$ 256\\
 Dec. & -1.3 $\pm$  1.2 &    32.2 $\pm$  1.1 &    11354 $\pm$   76 &    5639 $\pm$ 223\\
\hline
\end{tabular}
\end{center}
\end{table}

%Tucson Wind Speed Model
\begin{table}[htbp]
{\bf \caption{\label{tucson_coeffs}Tucson Wind Speed Model Coefficients}}
\begin{center}
\begin{tabular}{lcccc}
\hline
Month & A0 & A1 & A2 & A3\\
\hline
% Jan. & -3.8 $\pm$  1.1 &    36.7 $\pm$  1.0 &    11461 $\pm$   53 &    5902 $\pm$ 178\\
% Feb. & -0.6 $\pm$  0.9 &    36.1 $\pm$  0.9 &    11615 $\pm$   59 &    5278 $\pm$ 156\\
% Mar. & -0.9 $\pm$  1.0 &    36.1 $\pm$  1.0 &    11776 $\pm$   67 &    5477 $\pm$ 178\\
% Apr. &  0.2 $\pm$  1.2 &    31.3 $\pm$  1.1 &    11800 $\pm$   89 &    5409 $\pm$ 229\\
% May. &  2.6 $\pm$  1.0 &    23.0 $\pm$  0.9 &    11710 $\pm$  119 &    4898 $\pm$ 271\\
% Jun. &  5.3 $\pm$  0.6 &    15.5 $\pm$  0.7 &    12076 $\pm$  161 &    3755 $\pm$ 259\\
% Jul. &  4.2 $\pm$  0.5 &     7.1 $\pm$  0.5 &    11720 $\pm$  239 &    3801 $\pm$ 411\\
% Aug. &  4.3 $\pm$  0.3 &     8.2 $\pm$  0.4 &    11884 $\pm$  169 &    3173 $\pm$ 249\\
% Sep. &  4.9 $\pm$  0.5 &    17.0 $\pm$  0.6 &    12054 $\pm$  117 &    3524 $\pm$ 180\\
% Oct. &  3.3 $\pm$  0.7 &    23.6 $\pm$  0.7 &    11878 $\pm$  103 &    4458 $\pm$ 201\\
% Nov. & -2.1 $\pm$  1.2 &    32.5 $\pm$  1.1 &    11443 $\pm$   69 &    5676 $\pm$ 217\\
% Dec. & -2.3 $\pm$  1.1 &    34.5 $\pm$  1.0 &    11454 $\pm$   58 &    5620 $\pm$ 178\\
Jan. &   -1.1 $\pm$  0.7 &    34.1 $\pm$  0.6 &    11467 $\pm$   34 &    5549 $\pm$ 112\\
Feb. &    1.6 $\pm$  0.6 &    34.1 $\pm$  0.6 &    11620 $\pm$   40 &    4993 $\pm$ 104\\
Mar. &    1.4 $\pm$  0.7 &    34.0 $\pm$  0.6 &    11779 $\pm$   46 &    5163 $\pm$ 120\\
Apr. &    2.6 $\pm$  0.8 &    29.1 $\pm$  0.7 &    11804 $\pm$   63 &    5028 $\pm$ 155\\
May. &    4.6 $\pm$  0.7 &    21.2 $\pm$  0.6 &    11729 $\pm$   85 &    4465 $\pm$ 181\\
Jun. &    6.5 $\pm$  0.4 &    14.7 $\pm$  0.5 &    12107 $\pm$  109 &    3358 $\pm$ 167\\
Jul. &    5.4 $\pm$  0.2 &     6.4 $\pm$  0.3 &    11852 $\pm$  128 &    2933 $\pm$ 184\\
Aug. &    5.0 $\pm$  0.2 &     7.8 $\pm$  0.3 &    11955 $\pm$  101 &    2779 $\pm$ 141\\
Sep. &    5.7 $\pm$  0.3 &    16.3 $\pm$  0.4 &    12076 $\pm$   80 &    3282 $\pm$ 120\\
Oct. &    4.8 $\pm$  0.5 &    22.4 $\pm$  0.5 &    11889 $\pm$   76 &    4158 $\pm$ 144\\
Nov. &    0.5 $\pm$  0.8 &    30.1 $\pm$  0.8 &    11453 $\pm$   50 &    5297 $\pm$ 152\\
Dec. &    0.0 $\pm$  0.7 &    32.3 $\pm$  0.7 &    11462 $\pm$   41 &    5295 $\pm$ 123\\
\hline
\end{tabular}
\end{center}
\end{table}

%Flagstaff Wind Speed Model
\begin{table}[htbp]
{\bf \caption{\label{flagstaff_coeffs}Flagstaff Wind Speed Model Coefficients}}
\begin{center}
\begin{tabular}{lcccc}
\hline
Month& A0 & A1 & A2 & A3\\
\hline
 Jan. & -8.1 $\pm$  4.0 &    36.8 $\pm$  3.9 &    11486 $\pm$   63 &    6939 $\pm$ 528\\
 Feb. &  1.3 $\pm$  1.6 &    29.2 $\pm$  1.5 &    11404 $\pm$   55 &    5210 $\pm$ 244\\
 Mar. & -0.6 $\pm$  0.9 &    27.0 $\pm$  0.9 &    11487 $\pm$   29 &    5702 $\pm$ 155\\
 Apr. &  2.9 $\pm$  0.8 &    24.6 $\pm$  0.8 &    11280 $\pm$   34 &    5106 $\pm$ 151\\
 May. &  6.3 $\pm$  0.7 &    17.7 $\pm$  0.7 &    11569 $\pm$   68 &    4266 $\pm$ 192\\
 Jun. &  8.7 $\pm$  0.3 &    11.8 $\pm$  0.4 &    12084 $\pm$   80 &    2865 $\pm$ 123\\
 Jul. &  6.0 $\pm$  0.2 &     7.7 $\pm$  0.2 &    11847 $\pm$   68 &    2606 $\pm$ 101\\
 Aug. &  5.7 $\pm$  0.2 &    10.3 $\pm$  0.3 &    11967 $\pm$   61 &    2583 $\pm$  88\\
 Sep. &  8.0 $\pm$  0.4 &    16.2 $\pm$  0.5 &    12097 $\pm$   82 &    3133 $\pm$ 137\\
 Oct. &  4.5 $\pm$  0.8 &    19.8 $\pm$  0.7 &    11669 $\pm$   59 &    4587 $\pm$ 183\\
 Nov. & -0.2 $\pm$  1.7 &    29.0 $\pm$  1.6 &    11443 $\pm$   54 &    5484 $\pm$ 266\\
 Dec. & -2.9 $\pm$  2.8 &    31.7 $\pm$  2.7 &    11246 $\pm$   61 &    5969 $\pm$ 398\\
\hline
\end{tabular}
\end{center}
\end{table}

% Tenerife Wind Speed Model
\begin{table}[htbp]
{\bf \caption{\label{tenerife_coeffs}Tenerife Wind Speed Model Coefficients}}
\begin{center}
\begin{tabular}{lcccc}
\hline
Month& A0 & A1 & A2 & A3\\
\hline
Jan. &  2.5 $\pm$  0.9 &    20.3 $\pm$  0.9 &    12371 $\pm$  159 &    7964 $\pm$ 341\\
Feb. &  4.0 $\pm$  0.6 &    29.7 $\pm$  0.6 &    12102 $\pm$   82 &    6478 $\pm$ 149\\
Mar. &  1.7 $\pm$  1.1 &    27.5 $\pm$  1.0 &    12847 $\pm$  173 &    8041 $\pm$ 325\\
Apr. &  3.0 $\pm$  0.9 &    30.6 $\pm$  0.8 &    12332 $\pm$  120 &    6784 $\pm$ 216\\
May. &  4.6 $\pm$  0.5 &    24.9 $\pm$  0.5 &    12294 $\pm$   92 &    5646 $\pm$ 143\\
Jun. &  4.9 $\pm$  0.8 &    14.9 $\pm$  0.8 &    12711 $\pm$  282 &    5899 $\pm$ 420\\
Jul. &  6.7 $\pm$  0.4 &     8.1 $\pm$  0.6 &    12536 $\pm$  232 &    3760 $\pm$ 304\\
Aug. &  6.3 $\pm$  0.4 &     8.3 $\pm$  0.7 &    12137 $\pm$  226 &    3630 $\pm$ 296\\
Sep. &  6.0 $\pm$  0.4 &    11.5 $\pm$  0.6 &    12137 $\pm$  172 &    4341 $\pm$ 241\\
Oct. & -0.2 $\pm$  2.4 &    22.9 $\pm$  2.2 &    11545 $\pm$  218 &    8465 $\pm$ 661\\
Nov. &  4.5 $\pm$  0.7 &    16.2 $\pm$  0.7 &    12313 $\pm$  191 &    6579 $\pm$ 336\\
Dec. &  5.6 $\pm$  0.7 &    20.9 $\pm$  0.7 &    12074 $\pm$  139 &    6033 $\pm$ 238\\
\hline
\end{tabular}
\end{center}
\end{table}

%Antofagasta Wind Speed Model
\begin{table}[htbp]
{\bf \caption{\label{anto_coeffs}Antofagasta Wind Speed Model Coefficients}}
\begin{center}
\begin{tabular}{lcccc}
\hline
Month& A0 & A1 & A2 & A3\\
\hline
 Jan. &  3.8 $\pm$  0.3 &    13.4 $\pm$  0.3 &    12169 $\pm$   92 &    3357 $\pm$ 134\\
 Feb. &  3.5 $\pm$  0.3 &    14.1 $\pm$  0.3 &    12090 $\pm$   86 &    3598 $\pm$ 131\\
 Mar. &  4.0 $\pm$  0.3 &    17.3 $\pm$  0.4 &    12186 $\pm$   91 &    3540 $\pm$ 135\\
 Apr. &  4.0 $\pm$  0.4 &    25.3 $\pm$  0.5 &    12318 $\pm$   75 &    3727 $\pm$ 112\\
 May. &  2.7 $\pm$  0.4 &    29.7 $\pm$  0.4 &    12419 $\pm$   68 &    4367 $\pm$ 111\\
 Jun. &  3.0 $\pm$  0.5 &    30.5 $\pm$  0.5 &    12179 $\pm$   66 &    4298 $\pm$ 110\\
 Jul. &  3.1 $\pm$  0.6 &    32.5 $\pm$  0.6 &    11797 $\pm$   65 &    4148 $\pm$ 114\\
 Aug. &  2.7 $\pm$  0.5 &    31.1 $\pm$  0.5 &    11671 $\pm$   53 &    4274 $\pm$  98\\
 Sep. &  3.4 $\pm$  0.5 &    32.1 $\pm$  0.5 &    11785 $\pm$   59 &    3948 $\pm$  98\\
 Oct. &  3.7 $\pm$  0.4 &    29.1 $\pm$  0.4 &    12052 $\pm$   55 &    3785 $\pm$  86\\
 Nov. &  4.0 $\pm$  0.3 &    26.0 $\pm$  0.4 &    12192 $\pm$   60 &    3394 $\pm$  88\\
 Dec. &  4.0 $\pm$  0.3 &    21.1 $\pm$  0.4 &    12240 $\pm$   65 &    3131 $\pm$  91\\
\hline
\end{tabular}
\end{center}
\end{table}

\begin{figure}[htbp]
\centering\includegraphics[height=7.5cm]{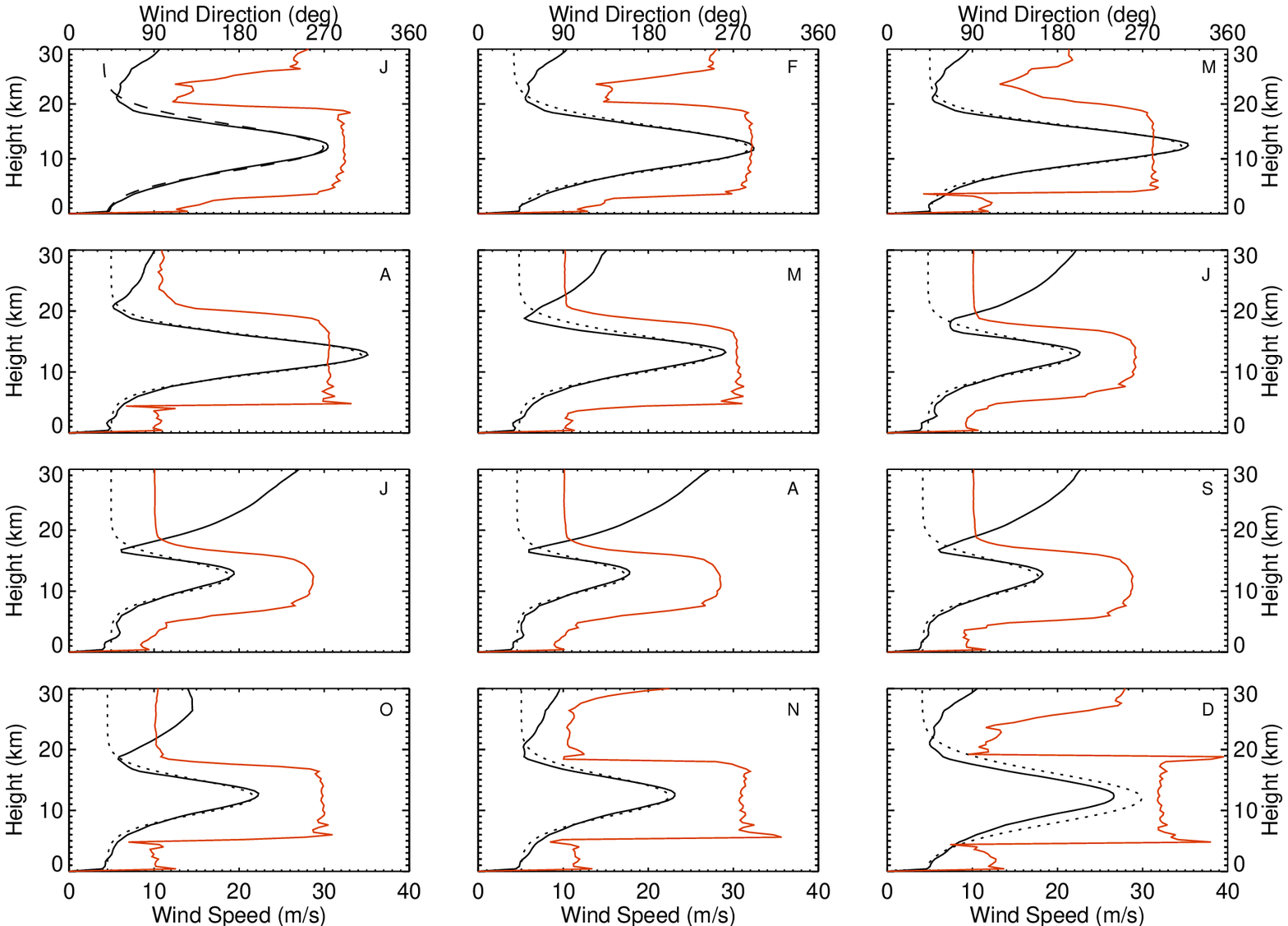}
\caption{\label{hilo_wind}  The wind speed and direction for Hilo, Hawaii.  Each of the graphs in this figure has the the measured wind speed as a function of height plotted with a solid black line and the wind speed computed from the model coefficients in Table \ref{hilo_coeffs} is shown as a black dashed line. The wind direction is plotted as a solid red line.  The wind speed is marked along the bottom of the graph, and the wind direction is marked on the top of the graph.  The first letter of the corresponding month is in the upper right corner. }
\end{figure}

\begin{figure}[htbp]
\centering\includegraphics[height=7.5cm]{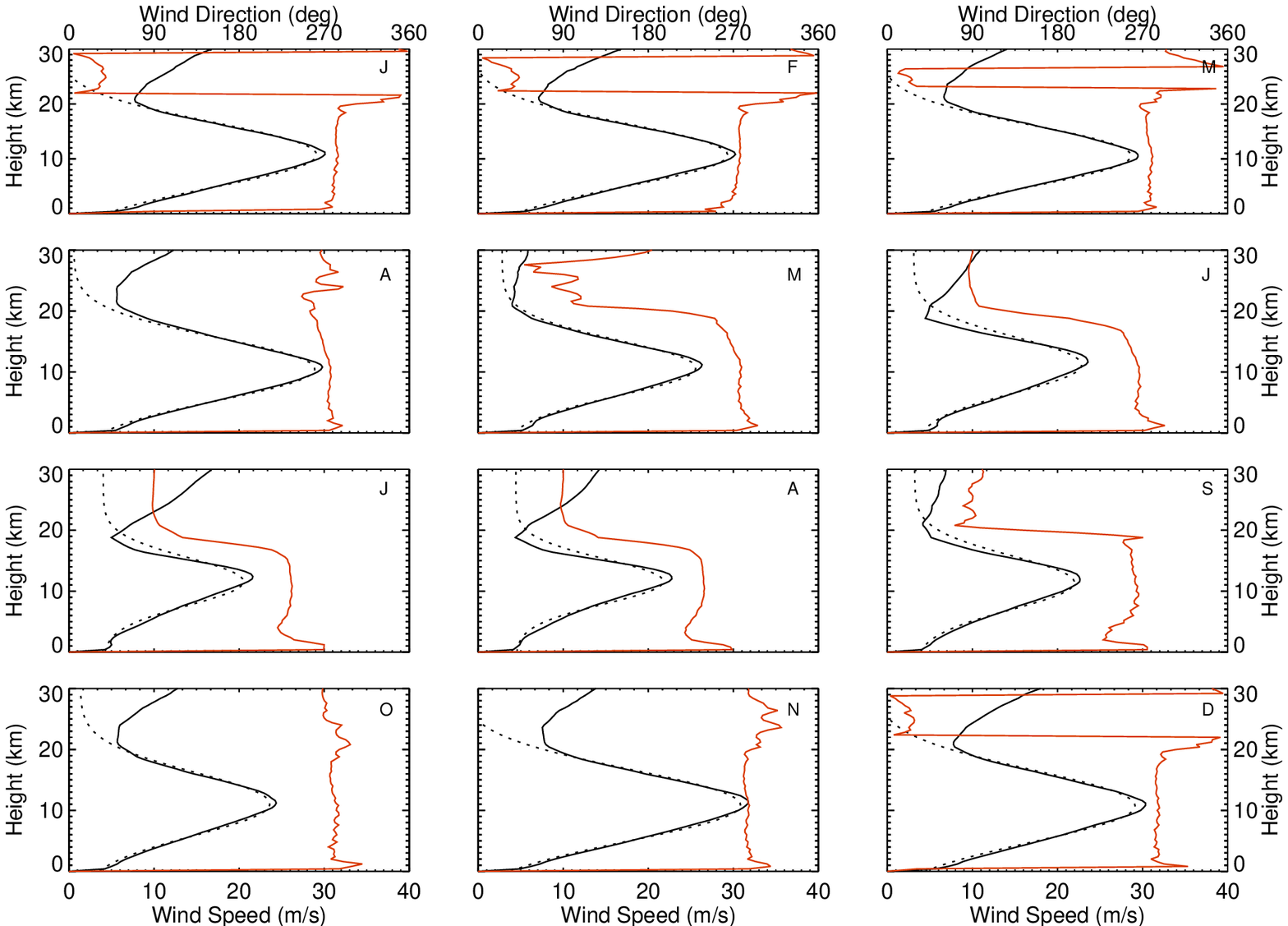} 
\caption{ \label{oakland_wind} The wind speed and direction for Oakland, California.  The layout of the figure is the same as in Fig. \ref{hilo_wind}, except the fitted wind model comes from Table \ref{oakland_coeffs}. }
\end{figure}

\begin{figure}[htbp]
\centering\includegraphics[height=7.5cm]{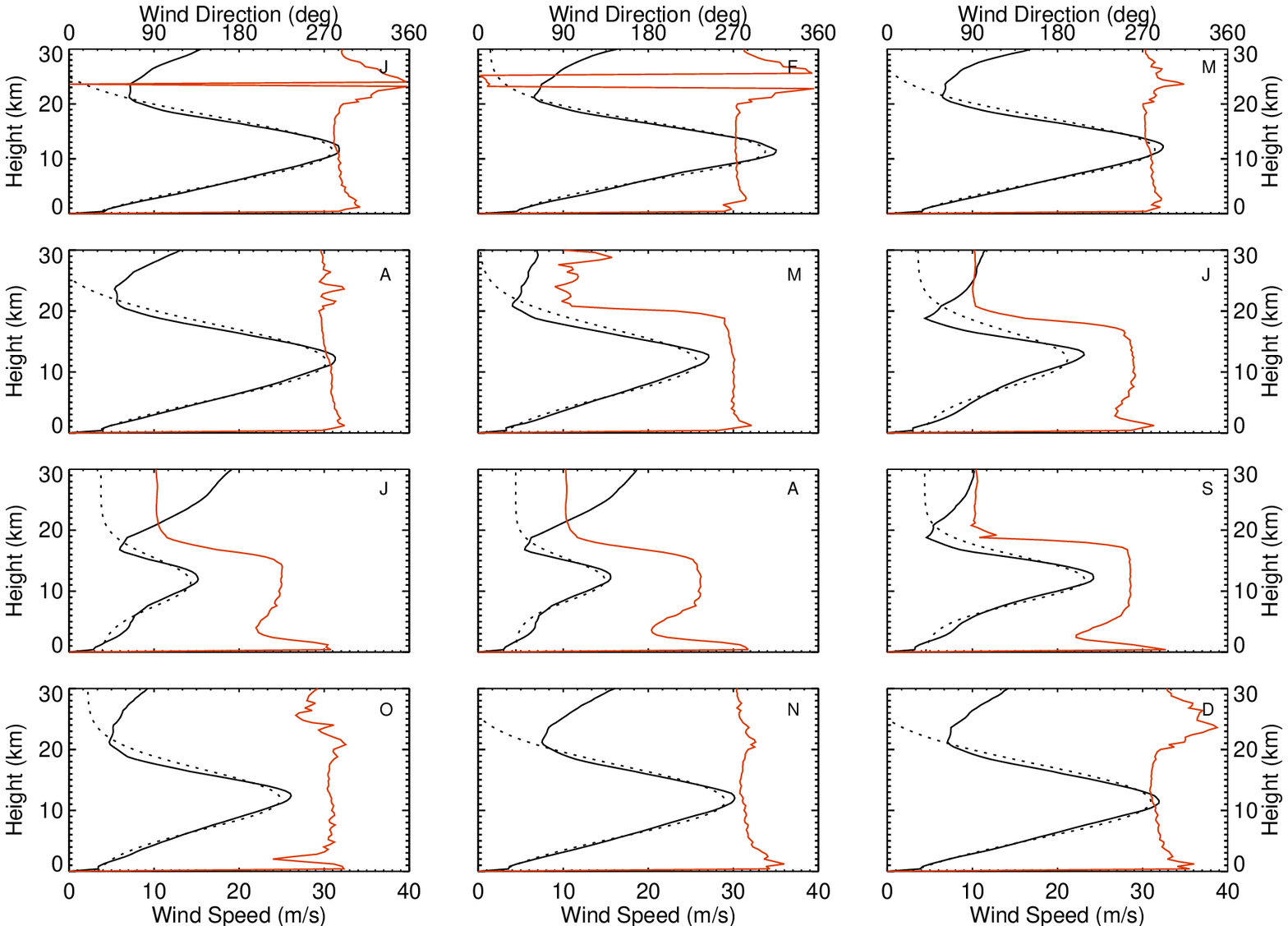}
\caption{\label{sandiego_wind} The wind speed and direction for San Diego, California.  The layout of the figure is the same as in Fig. \ref{hilo_wind}, except the fitted wind model comes from Table \ref{sandiego_coeffs}.}
\end{figure}

\begin{figure}[htbp]
\centering\includegraphics[height=7.5cm]{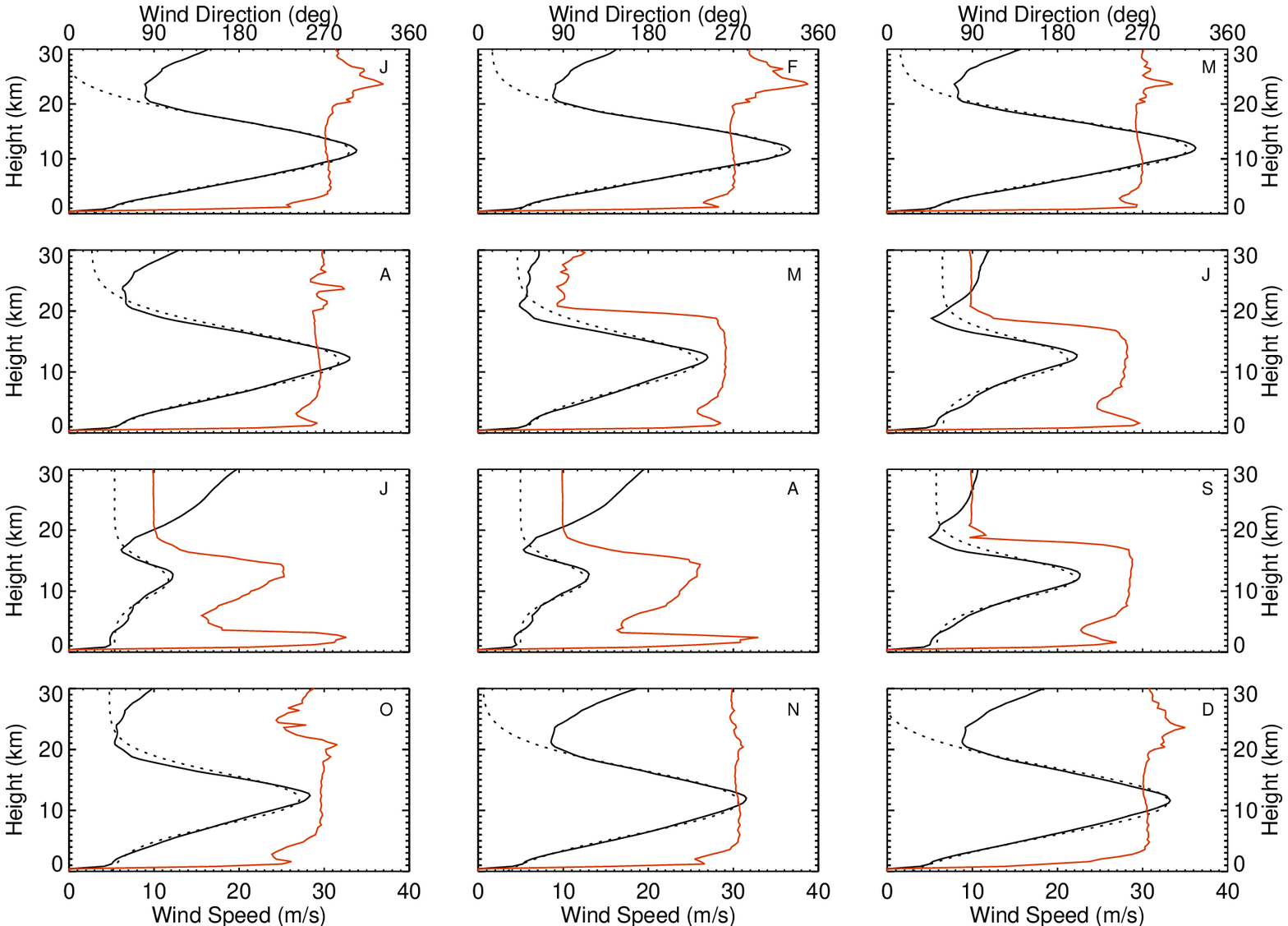} 
\caption{\label{tucson_wind}  The wind speed and direction for Tucson, Arizona. The 
layout of the figure is the same as in Fig. \ref{hilo_wind}, except the fitted wind model comes from Table \ref{tucson_coeffs}.  }
\end{figure}

\begin{figure}[htbp]
\centering\includegraphics[height=7.5cm]{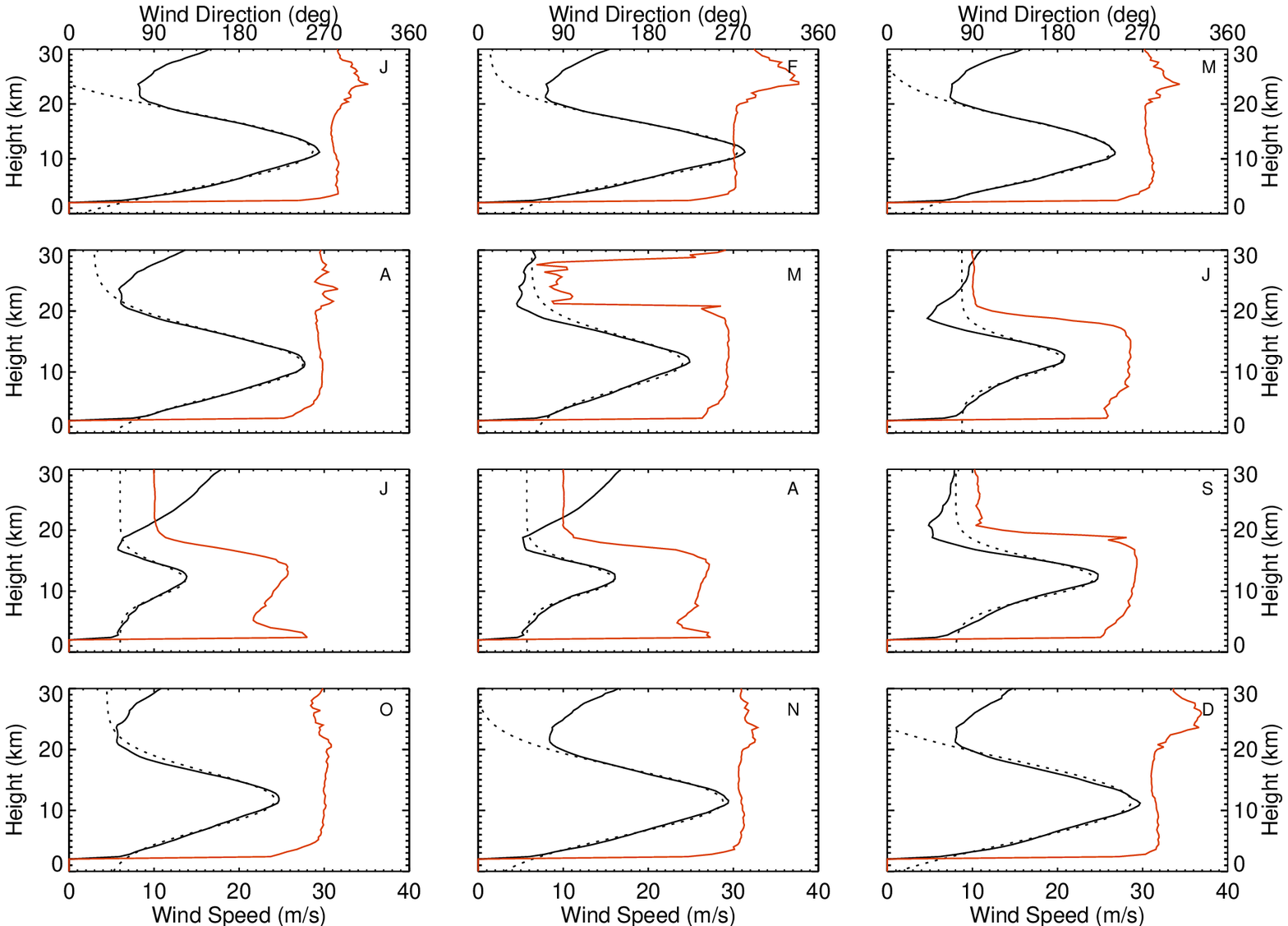}
\caption{\label{flagstaff_wind}  The wind speed and direction for Flagstaff, Arizona. 
The layout of the figure is the same as in Fig. \ref{hilo_wind}, except the fitted wind model comes from Table \ref{flagstaff_coeffs}. }
\end{figure}

\begin{figure}[htbp]
\centering\includegraphics[height=7.5cm]{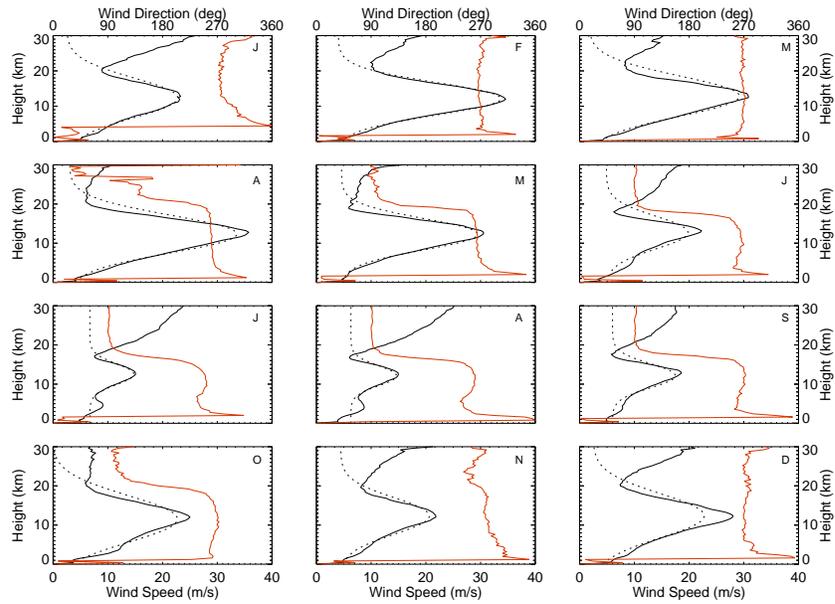} 
\caption{\label{tenerife_wind}  The wind speed and direction for Tenerife, Canary Islands, Spain, which is useful for the observatories located in the Canary Islands. The layout of the figure is the same as in Fig. \ref{hilo_wind}, except the fitted wind model comes from Table \ref{tenerife_coeffs}.     }
\end{figure}

\begin{figure}[htbp]
\centering\includegraphics[height=7.5cm]{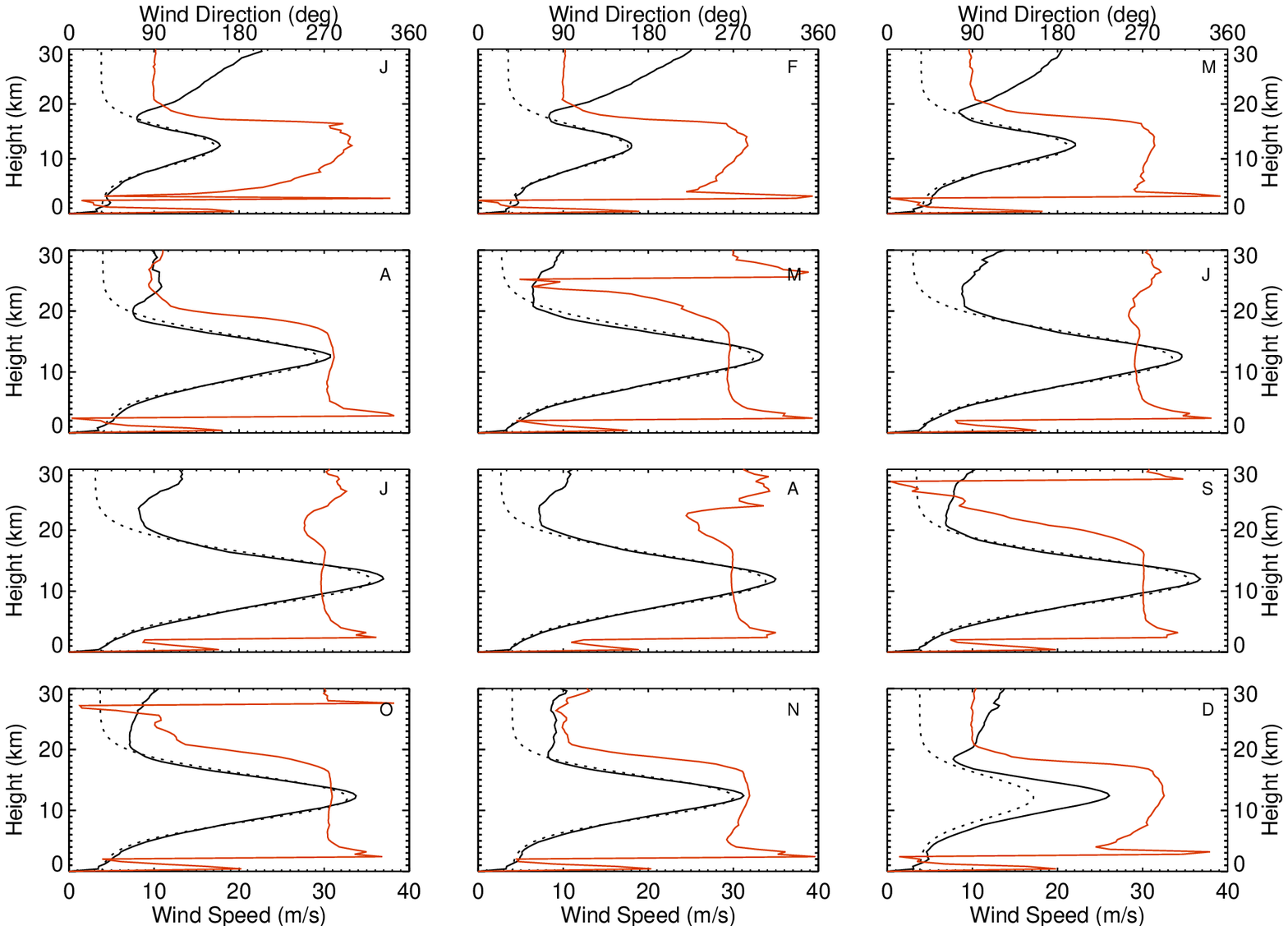} 
\caption{\label{anto_wind}  The wind speed and direction for Antofagasta, Chile. The layout of the figure is the same as in Fig. \ref{hilo_wind}, except the fitted wind model comes from Table \ref{anto_coeffs}. }
\end{figure}

The results are shown in Figs. \ref{hilo_wind}--\ref{anto_wind} respectively.    Each figure is composed of 12 graphs corresponding to the 12 months of the year.  The first letter of the corresponding month is in the upper right corner.  Each graph has the the wind speed plotted with a solid line and the fitted wind speed computed from the coefficients in Tables \ref{hilo_coeffs}--\ref{anto_coeffs} is plotted as a dashed line. The wind direction is plotted as a solid red line.  The wind speed is marked along the bottom of the graph, while the wind direction is marked on the top of the graph. The wind directions often show sharp spikes at the edges of jets. Those are an artificial result of the averaging of directions and can be ignored.

For both hemispheres, the tropospheric winds increase in the winter months and decrease in the summer months. Stratospheric winds increase in the summer months, the opposite of the tropospheric winds.  During the summer months, Tenerife has a low level ($\approx5$ km) jet form. This is most obvious in July and August, but first appears in June and persists to October.   In the Greenwood model, $A_0$ is the low altitude wind speed and that is true for our model for Hawaii.  It is not always true for other sites, as these values are sometimes negative. The reason for this is that the wind profiles for those sites has a slightly different shape than the Hawaiian wind profile.  This can be seen in Figs.  \ref{oakland_wind} and\ref{flagstaff_wind}.  

Greenwood \cite{greenwood1977} only provides a model for the wind speed, though \cite{greenwood1975} does show the annual mean wind direction for Hilo and Lihue.  As shown by Figs. \ref{hilo_wind}-\ref{anto_wind} the wind direction is not an easily modeled function.  It can most easily be modeled as a square wave.  Since stability of flows does depend upon the amount of shear present, an analysis that computed the wind shear at the upper and lower boundaries would tend to overestimate the instability.    Looking at Figs. \ref{hilo_wind}--\ref{anto_wind}, the wind direction does not change instantaneously, but it does change very quickly, often changing direction by $180^\circ$ in a km.  Some regions of slowly changing direction, most commonly above 25 km,  are not very well modeled by a square wave. A square wave model will suffice for modeling where phase screens at only a few altitudes are used to model the distribution of turbulence. 

Note that the reversal of wind direction at altitudes above 20 km is a well known feature of the zonal wind flow. It is due to a higher altitude jet whose center is north  of the equator in winter, and south of the equator in summer.  There are also northern and southern jets at higher altitudes which move in the same direction as the jets in the lower stratosphere. These are seasonal and impact the wind speed and directions in our data \cite{newell1972}.

\section{Richardson Number}\label{richardson}

The gradient Richardson number is a dimensionless ratio, $R_i$, related to the buoyant production or consumption of turbulence divided by the shear production of turbulence \cite{kundu1990}. It is defined as 

\eq
\label{eq_richardson}
R_i = \frac{g\frac{\partial\ln\Theta}{\partial~z}}{\left (\frac{\partial u}{\partial z}\right)^2 + \left(\frac{\partial v}{\partial z}\right)^2},
\en

where $g$ is the acceleration of gravity, $\Theta$ is the potential temperature, $z$ is the altitude, $u$ and $v$ are the horizontal components of the wind vector.  Normally a gradient Richardson number with a value less than 0.25 is considered to be turbulent and a value greater than 0.25 to be non-turbulent.  It is used to indicate dynamic stability and the formation of turbulence.

In order to see at which altitudes turbulence forms, we used data from the radiosondes to compute the monthly binned Richardson number for each site in Table \ref{station_table} using Eq. \ref{eq_richardson} at each of the fixed grid points used in \S \ref{data_analysis}.  The grid heights points are every 400m and some of the finer atmospheric layers may not be captured with our analysis.  The derivatives were computed using a three-point Lagrangian interpolation routine.  

\begin{figure}[htbp]
\centering\includegraphics[height=7.5cm]{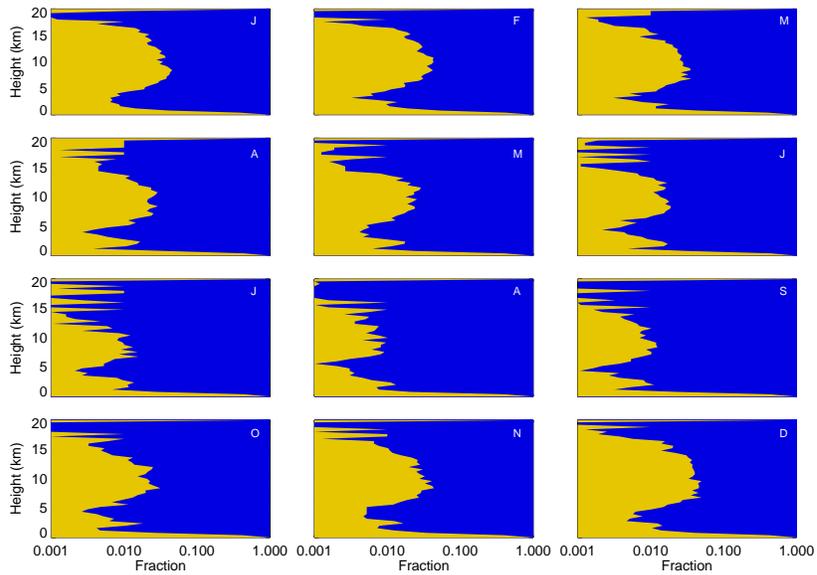} 
\caption{\label{hilo_rich} 
The binned Richardson number for Hilo, Hawaii The figure shows the fraction of Richardson numbers falling into two bins.  Yellow signifies turbulent conditions ($R_i < 0.25$) and blue signifies non-turbulent conditions ($R_i \ge 0.25$).}
\end{figure}

\begin{figure}[htbp]
\centering\includegraphics[height=7.5cm]{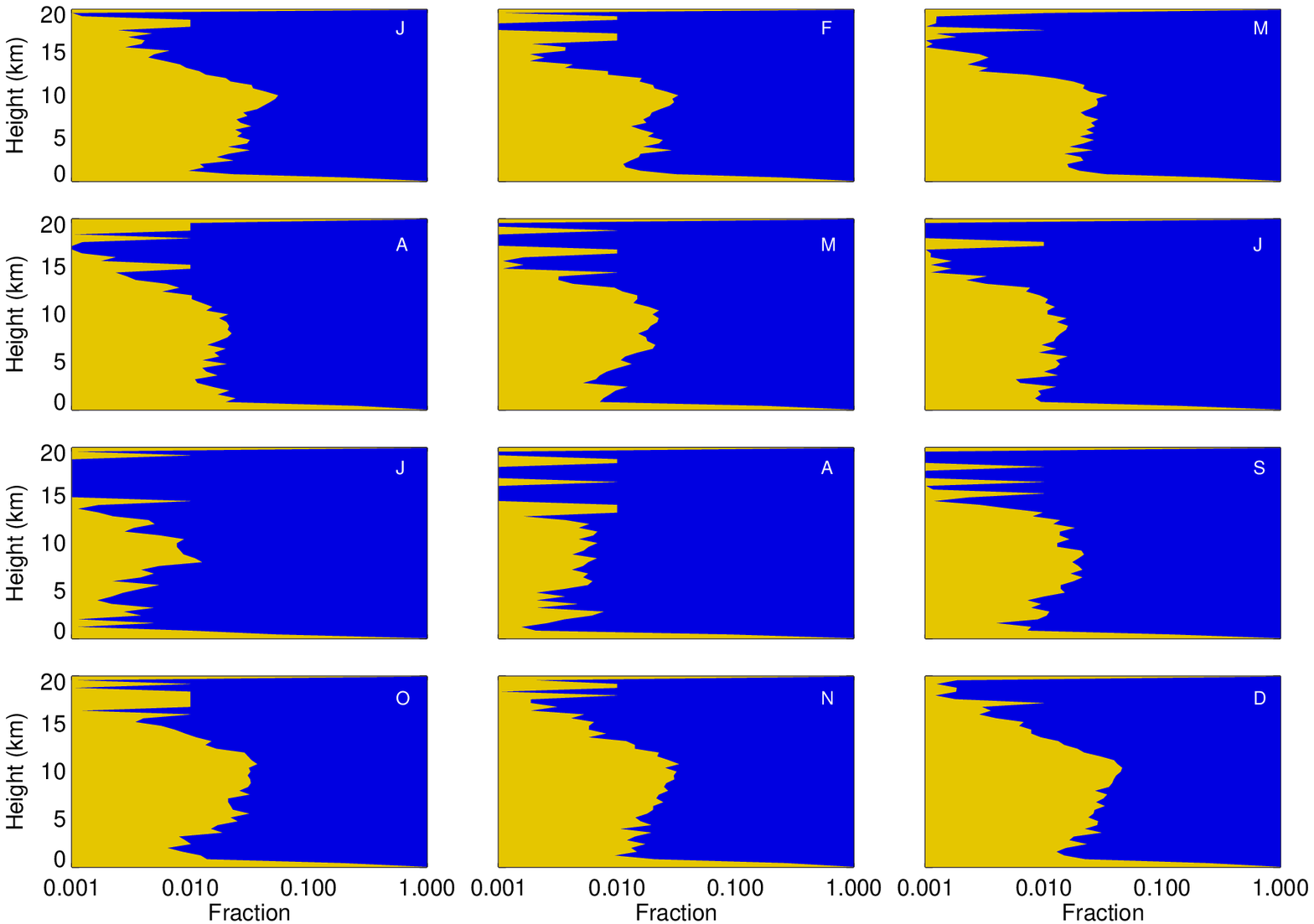} 
\caption{\label{oakland_rich}   The binned Richardson number for Oakland, California.  Yellow signifies turbulent conditions ($R_i < 0.25$) and blue signifies non-turbulent conditions ($R_i \ge 0.25$).}
\end{figure}

\begin{figure}[htbp]
\centering\includegraphics[height=7.5cm]{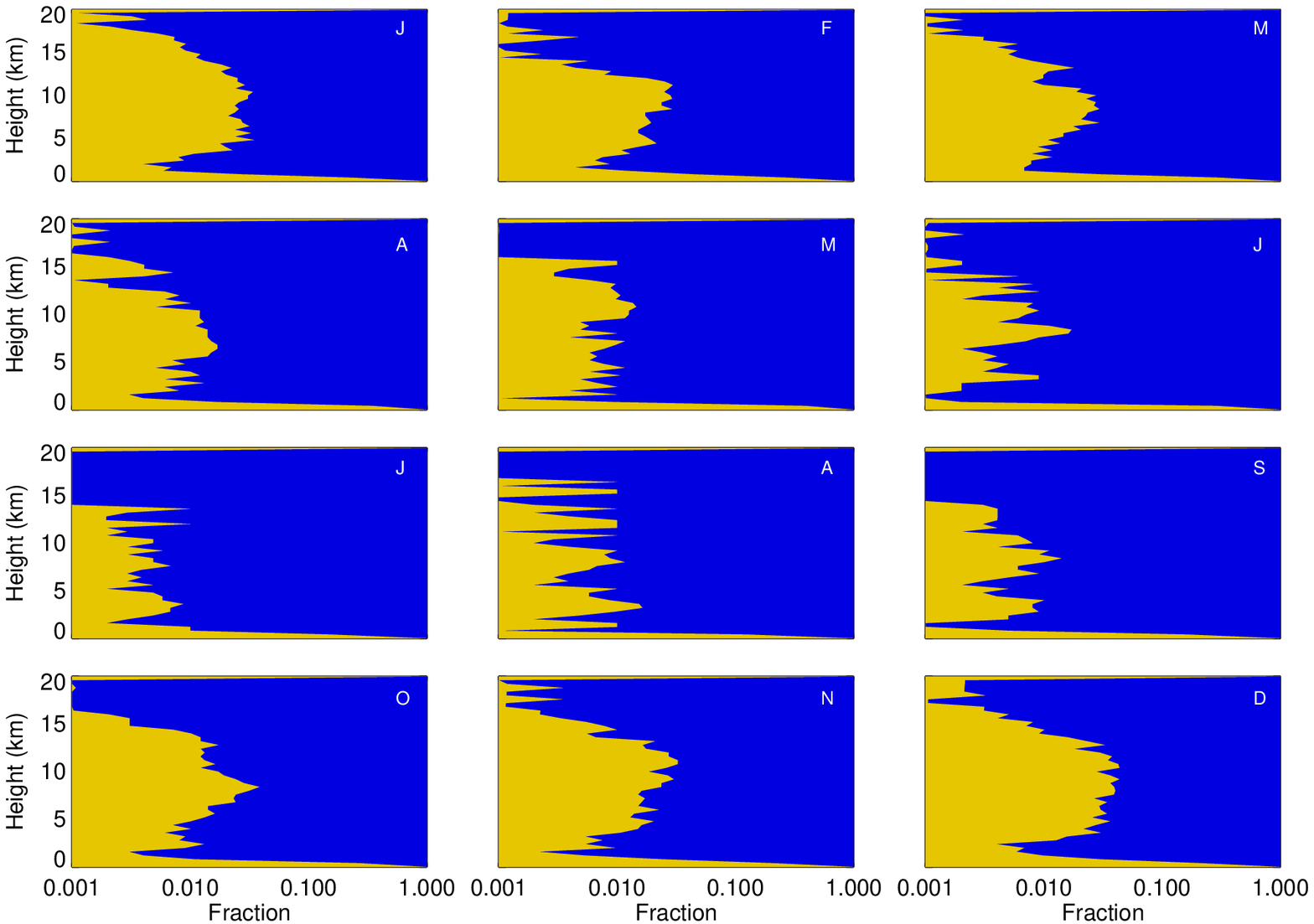} 
\caption{\label{sandiego_rich}  The binned Richardson number for San Diego, California.   Yellow signifies turbulent conditions ($R_i < 0.25$) and blue signifies non-turbulent conditions ($R_i \ge 0.25$). }
\end{figure}

\begin{figure}[htbp]
\centering\includegraphics[height=7.5cm]{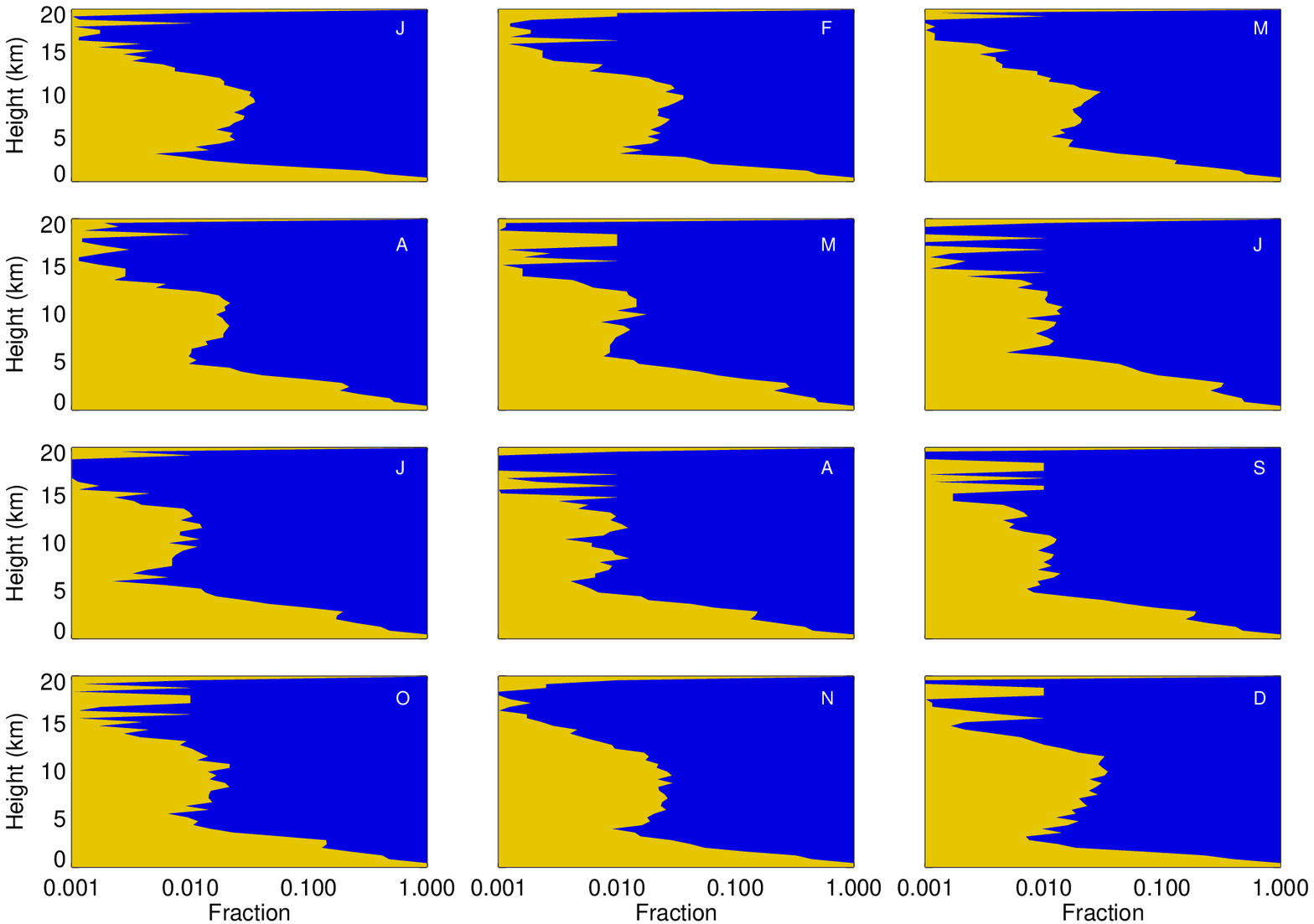} 
\caption{ \label{tucson_rich}   The binned Richardson number for Tucson, Arizona.  Yellow signifies turbulent conditions ($R_i < 0.25$) and blue signifies non-turbulent conditions ($R_i \ge 0.25$). }
\end{figure}

\begin{figure}[htbp]
\centering\includegraphics[height=7.5cm]{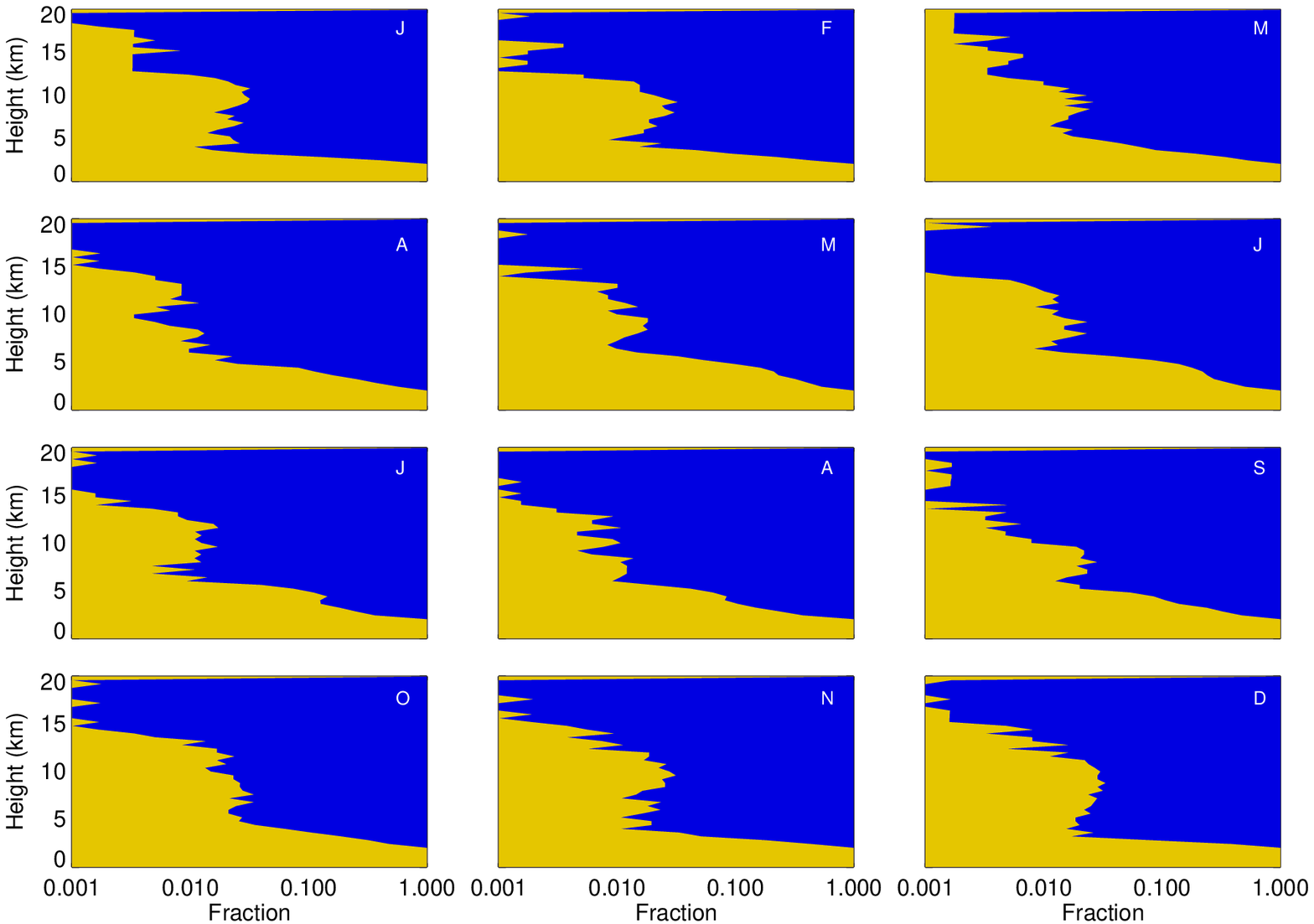} 
\caption{ \label{flagstaff_rich}   The binned Richardson number for Flagstaff, Arizona.  Yellow signifies turbulent conditions ($R_i < 0.25$) and blue signifies non-turbulent conditions ($R_i \ge 0.25$). }
\end{figure}

\begin{figure}[htbp]
\centering\includegraphics[height=7.5cm]{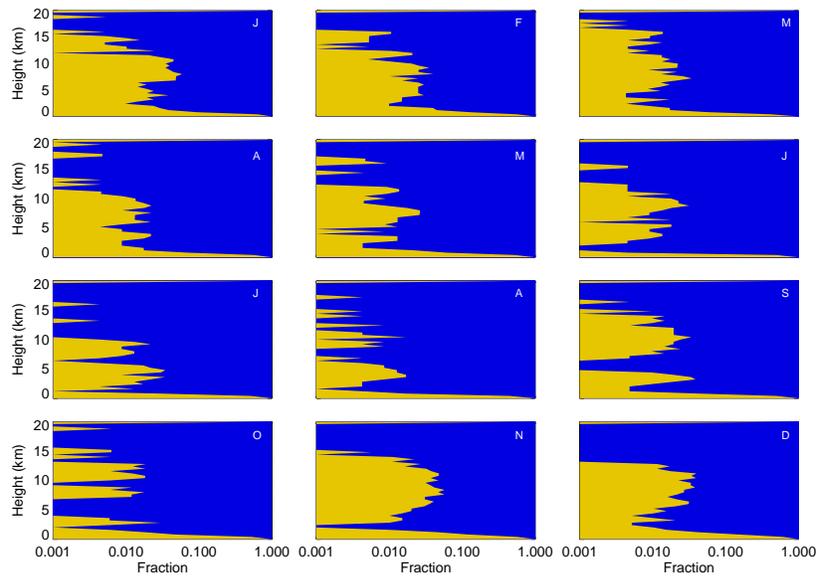} 
\caption{\label{tenerife_rich}   The binned Richardson number for Tenerife, Canary Islands, Spain.  Yellow signifies turbulent conditions ($R_i < 0.25$) and blue signifies non-turbulent conditions ($R_i \ge 0.25$). }
\end{figure}

\begin{figure}[htbp]
\centering\includegraphics[height=7.5cm]{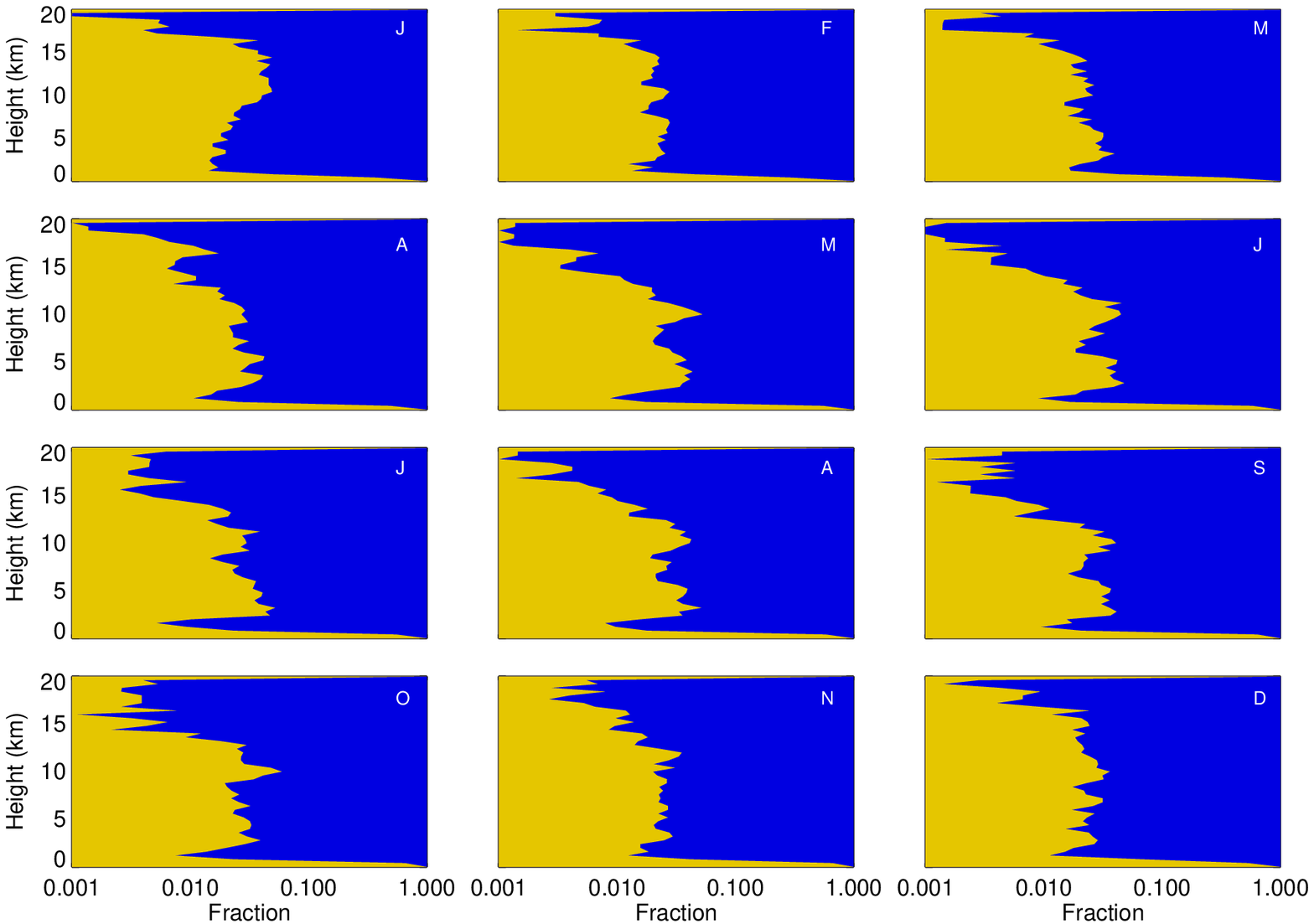} 
\caption{\label{anto_rich}  The binned Richardson number for Antofagasta, Chile.  Yellow signifies turbulent conditions ($R_i < 0.25$) and blue signifies non-turbulent conditions ($R_i \ge 0.25$).}
\end{figure}

The computed Richardson numbers were then binned into two bins: $R_i < 0.25$, and  $R_i \geq 0.25$. 
For a given site, at each altitude the numbers in each bin were normalized by the total number of data points at that altitude.  This gives the Richardson number as a function of altitude.  These are shown in Figs. \ref{hilo_rich}-\ref{anto_rich}. As expected, the Richardson numbers above 20 km are very high and have little impact on astronomical seeing; as a result we only plotted the values below 20 km. It is important to note that these values are for radiosondes launched from sites at lower altitudes than astronomical observatories.  The highest Richardson number values occur at these low altitudes and are not a concern to most astronomical observatories.

\

Examining the individual months reveals that the figures are not as smooth as the wind speed figures, but instead have many fine layers.  The layers are significant, even if they do not seem to explicitly show turbulence (ie their value of $R_i$ is > 0.25). Because the $R_i$ calculation has a height increment of 400m, and because the precise positions and altitudes of jets vary over periods of a day or less, the effective values of the velocity derivatives in Equation \ref{eq_richardson} are lower in our results than they might be at any given time. This averaging leads to relatively large calculated $R_i$ values. It also means that our data will not show many instances of $R_i < 0.25$.  But where is does occur it is significant. Moreover, any drop in the plotted $R_i$ number indicates a persistent region of instability,  and it is likely that some turbulence is present there for extended periods,

Since low gradient Richardson numbers indicate the presence of atmospheric turbulence, they can be an indicator of seeing. In Fig. \ref{hilo_rich}, the winter months for Hawaii have the highest proportion of Richardson numbers lower than 0.25 (yellow values), and the summer months have the lowest. In other words the winter months have more turbulent conditions than the summer months. Studies of seeing from Mauna Kea show that this seasonal behavior of seeing correlates with this very well \cite{bely1987,seigar2002, skidmore2009}.  There is relatively little seasonal variation in for Antofagasta, Chile (Fig. \ref{anto_rich}) and this correlates well with seeing measurements carried at nearby Cerro Armazones \cite{skidmore2009} which show no clear evidence for any seasonal dependency.  The trend in Fig. \ref{tenerife_rich} agrees with seeing measurements for  La Palma, Canary Islands, showing that seeing is best May-September \cite{munoz1998, wilson1999}.  Flagstaff, Arizona (Fig. \ref{flagstaff_rich} has more turbulent values than Oakland, California, which is agreement with site comparison done between Anderson Mesa (outside of Flagstaff) and Lick Observatory (near Oakland) \cite{walters1997}.  This shows that Richardson numbers computed from radiosonde data can be used to differentiate between astronomical sites at a macro level which can be of use for narrowing site selection for new telescopes.  It is important to remember that the Richardson numbers computed from radiosonde data do not include local effects such as the turbulent surface boundary layer that can have a profound impact on seeing.  Small differences in altitudes and location can alter the observed seeing significantly \cite{walters1997}.  

While the Richardson number does not indicate the strength of turbulence, knowing that there are persistent turbulent layers may be useful to site planners.  For example, knowing roughly where and when one might expect upper layers of turbulence to be present could be useful for measurements, such as laser communications, which are strongly influenced by scintillation. On the other hand, we have seen that some layers closer to the ground have $R_i$<0.25. These must also be persistent to show up over the averaging. Knowing that these layers exist is useful for large telescopes
that plan to compensate for near ground effects.

\section{Other Wind Models}\label{theothers}

There have been a few other studies of wind speed for different sites, though they have been done for only a limited number of sites and with different methodology.  Sarazin \& Tokovinin \cite{sarazin2002} suggested that the wind speed at the 200 mbar pressure isobar (\vtwo)~is a good indication of atmospheric turbulence at a given site (\vtwo~is well monitored for meteorology).    The advantage of this is that there are models of \vtwo~for the entire planet, while there are only a fixed number of radiosondes launched each day.  Several groups have used this idea to characterize and compare several astronomical sites \cite{chueca2004, carrasco2005, garcia-lorenzo2005} including several sites which are included in this study.    During the year, the altitude at which \vtwo~occurs varies, this makes it difficult to use the values in Eq. \ref{greenwood_fit}. It is useful for site comparisons, but it supplements rather than replaces the knowledge of the vertical wind profile. 

Garc\'{i}a-Lorenzo et al. \cite{garcia-lorenzo2005} studied \vtwo~for Paranal and La Silla in Chile, San Pedro M\'{a}rtir in Mexico, Mauna Kea in Hawaii and La Palma in the Canary Islands.  The last two observatories are also characterized in this paper. We show a seasonal variation in $A_1$ in Tables \ref{hilo_coeffs} and \ref{tenerife_coeffs} which has the highest tropospheric wind speed in the spring and the lowest tropospheric winds in the summer. Garc\'{i}a-Lorenzo et al. showed that the \vtwo wind speeds at Mauna Kea and La Palma have this same seasonal behavior.   

\section{Conclusions}  

We have presented upper-level wind models computed from archival radiosonde data suitable for the major astronomical observatories in the United States, Chile and the Canary Islands.    We find that the commonly used Greenwood model  is not suitable for sites other than Hawaii. In addition it produced a single wind profile for the entire year, when it is more correct to use different wind models over the course of the year. We find that, as may be expected from geophysical fluid dynamics, that sites with similar latitudes share similar wind profiles. In addition we have computed Richardson number profiles from the same datasets. Those results indicate the presence of turbulence at different altitudes with seasonal variations that seem to agree with variations in seeing.  These models and results will be of use to modelers of atmospheric turbulence and instrument developers.

%%%%%%%%%%%%%%%%%%%%%%%%%%%%%%%%%%%%%%%%%%%%%%%%%%%%%%%%%%%%%
\section*{Acknowledgments}

This research was funded by the Air Force Office of Scientific Research and by the Air Force Research Laboratory's Directed Energy Directorate. A portion of the research in this paper was carried out at the Jet Propulsion Laboratory, California Institute of Technology, under a contract with the National Aeronautics and Space Administration. Radiosonde data were obtained from the Wyoming Weather Web, maintained by the Department of Atmospheric Science of the University of Wyoming.


\begin{thebibliography}{}

\bibitem[1]{hufnagel1974}
        R.E. Hufnagel,``Variations of Atmospheric Turbulence,'' in \emph{Topical Meeting on 
Optical Propagation through Turbulence}, (Optical Society of America,  1974), paper WA1-1.  

\bibitem[2]{beland1993}
       R.R. Beland, ``Propagation through Atmospheric Optical Turbulence,'' in \textit{The Infrared \& Electro-Optical Systems Handbook}, Vol. 2, F.G. Smith, ed. (SPIE, Bellingham, 1993), 158-232.  

\bibitem[3]{dewan1993}
       E.M. Dewan, R.E., Good, R. Beland, \& J. Brown, ``A Model for $C_n^2$ (Optical Turbulence) Profiles Using Radiosonde Data,'' PL-TR-93-2043, (Phillips Laboratory, Albuquerque 1993).

\bibitem[4]{taylor1938}
      G.I. Taylor, ``The Spectrum of Turbulence,'' Proc. R. Soc. London, Ser. A 
\textbf{164}, 476--490 (1938).

\bibitem[5]{bufton1973}
       J.L. Bufton,  ``Comparison of vertical profile turbulence structure with stellar observations,'' \ao  \textbf{12}, 1785--1793 (1973).

\bibitem[6]{parenti1994}
       R.R. Parenti,  \& R.J. Sasiela, ``Laser guide-star systems for astronomical applications,'' \josaa  \textbf{11}, 288--309 (1994).

\bibitem[7]{andrews1998}
       L.C. Andrews, \& R.L. Phillips, \textit{Laser Beam Propagation Through Random Media} (SPIE, 1998).

\bibitem[8]{tyson1998}
       R.K. Tyson, \textit{Principles of Adaptive Optics,} (Academic, San Diego, 1998).

\bibitem[9]{greenwood1977}
       Greenwood, D.P. ``Bandwidth specification for adaptive optics systems,'' 1977, \josa  \textbf{67,} 390.

\bibitem[10]{greenwood1975}
       D.P. Greenwood, \& D.L. Fried, ``Power Spectra Requirements for Wavefront-Compensative Systems,'' RADC-TR-75-227, (Rome Air Development Center, Rome, 1975).

\bibitem[11]{hardy1998}
        J.W. Hardy, \textit{Adaptive optics for astronomical telescopes},  (Oxford Univ., New York, 1998). 

\bibitem[12]{kundu1990}
  P.K. Kundu, \textit{Fluid Mechanics}, (Academic, San Diego, 1990).

\bibitem[13]{fesen2006}
       R.A. Fesen, ``A high-altitude station-keeping astronomical platform,'' \pspie  \textbf{6267}, 62670T, (2006).

\bibitem[14]{newell1972}
       R.E. Newell, J.W. Kidson, D.G. Vincent, \& G.J. Boer, \textit{The general circulation of the tropical atmosphere and interactions with extratropical latitudes,} Vol 1, (MIT, Cambridge, 1972).

\bibitem[15]{bely1987}
       P. Bely, ``Weather and seeing on Mauna Kea,'' Pub. Astro. Soc. Pacific  \textbf{99}, 560--570 (1987).

\bibitem[16]{seigar2002}
       M.S. Seigar, A.J. Adamson, N.P. Rees, T.G. Hawarden, M.J. Currie, \& T.C. Chuter, ``Seeing statistics at the upgraded 3.8m UK Infrared Telescope'', \pspie  \textbf{4844}, 366, (2002).

\bibitem[17]{skidmore2009}
         W. Skidmore, S. Els, T. Travouillon, R. Riddle, M. Sch\"ock, E. Bustos, J. Seguel, \& D. Walker, ``Thirty Meter Telescope Site Testing V: Seeing and Isoplanatic Angle,'' Pub. Astro. Soc. Pacific  \textbf{121}, 1151-1166, (2009).

\bibitem[18]{munoz1998}
      C. Mu\~{n}oz-Tu\~{n}\'{o}n, A.M. Varela,  \& T. Mahoney, ``Homogeneity of image quality at the Roque de los Muchachos Observatory,'' New Astronomy Reviews  \textbf{42}, 409--416, (1998).

\bibitem[19]{wilson1999}
       R.W. Wilson, N. O'Mahony, C. Packham, \& M. Azzaro, ``The seeing at the William Herschel Telescope,'' 1999, Mon. Not. Royal Astro. Soc. \textbf{309}, 379--387 (1999).

\bibitem[20]{walters1997}
       D.L. Walters,   \& L.W. Bradford, ``Measurements of $r_0$ and $\tau_0$ : two decades and 18 sites,'' \ao \textbf{36}, 7876--7886 (1997).

\bibitem[21]{sarazin2002}
         M. Sarazin, \& A. Tokovinin,  ``The statistics of isoplanatic angle and adaptive optics time constant derived from DIMM data,'', in \textit{Beyond  conventional adaptive optics : a conference devoted to the development  of adaptive optics for extremely large telescopes}.  E. Vernet, R.  Ragazzoni, S. Esposito, \& N. Hubin,eds (European Southern Observatory, Garching, 2002), 321--328.

\bibitem[22]{chueca2004}
       S. Chueca, B. Garc\'{i}a-Lorenzo, E.G. Mendiz\'{a}bal, T. Varela, J.J. Fuensalida, \& C. Mu\~{n}onz-Tu\~{n}\'{o}n, ``Input parameters of the HV model above Canarian observatories.'' \pspie \textbf{5237}, 159--166 (2004).

\bibitem[23]{carrasco2005}
       E. Carrasco, R. Avila, \& A. Carrami\~{n}ana, ``High-altitude wind velocity at Sierra Negra and San Pedro M\'{a}rtir,'' Pub. Astro. Soc. Pacific \textbf{117}, 104--110 (2005).

\bibitem[24]{garcia-lorenzo2005}
       B. Garc\'{i}a-Lorenzo, J.J. Fuensalida, C. Mu\~{n}onz-Tu\~{n}\'{o}n, \& E. Mendizabal, ``Astronomical site ranking based on tropospheric wind statistics,'' Mon. Not. Royal Astro. Soc. \textbf{356}, 849--858 (2005).



\end{thebibliography}
\end{document}